%% file: CHI26-1528.tex
\documentclass[sigconf]{acmart}

\usepackage{enumitem}
\usepackage{xspace}
\usepackage{caption} 
\usepackage{subcaption} 
\usepackage{multirow}

\usepackage{amsthm}%
\usepackage[title]{appendix}%
\usepackage{xcolor}
\usepackage{textcomp}%
\usepackage{manyfoot}
\usepackage{algorithm}%
\usepackage{listings}%
\usepackage{ifthen}
\usepackage{subcaption}
\usepackage{makecell}
\usepackage{colortbl}
\usepackage{soul}
\usepackage{verbatim}
\usepackage{makecell}
\usepackage{multirow}
\usepackage{ifthen}
\usepackage{colortbl}

\usepackage{rotating} 

\newboolean{showcomments}
\setboolean{showcomments}{false} 

\newboolean{useselfeff}
\setboolean{useselfeff}{true}

\newboolean{usehelpfulnessmoderator}
\setboolean{usehelpfulnessmoderator}{false}

\newcommand{\todo}{\ifthenelse{\boolean{showcomments}}{\textcolor{red}{TODO}\xspace}{}}
\newcommand{\ryan}[1]{\ifthenelse{\boolean{showcomments}}{\textcolor{cyan}{[#1 —ryan]}}{}}
\newcommand{\ifdita}[1]{\ifthenelse{\boolean{showcomments}}{\textcolor{brown}{[#1 —ifdita]}}{}}
\newcommand{\raj}[1]{\ifthenelse{\boolean{showcomments}}{\textcolor{orange}{[#1 —raj]}}{}}
\newcommand{\cut}[1]{\ifthenelse{\boolean{showcomments}}{\textcolor{red}{#1}}{}}
\newcommand{\emma}[1]{\ifthenelse{\boolean{showcomments}}{\textcolor{purple}{[#1 —emma]}}{}}
\newcommand{\diyi}[1]{\ifthenelse{\boolean{showcomments}}{\textcolor{blue}{[#1 —diyi]}}{}}
\newcommand{\juan}[1]{\ifthenelse{\boolean{showcomments}}{\textcolor{green}{[#1 —juan]}}{}}

\newcommand{\system}{\textsf{\scshape CARE}\xspace}

\definecolor{orange}{HTML}{C54524}
\definecolor{teal}{HTML}{65D1BF}
\definecolor{purple}{HTML}{8841E2}

\copyrightyear{2026}
\acmYear{2026}
\setcopyright{cc}
\setcctype{by}
\acmConference[CHI '26]{Proceedings of the 2026 CHI Conference on Human Factors in Computing Systems}{April 13--17, 2026}{Barcelona, Spain}
\acmBooktitle{Proceedings of the 2026 CHI Conference on Human Factors in Computing Systems (CHI '26), April 13--17, 2026, Barcelona, Spain}
\acmPrice{}
\acmDOI{10.1145/3772318.3791821}
\acmISBN{979-8-4007-2278-3/2026/04}

\begin{document}


\title[A Randomized Study with 90+ Novice Counselors]{Can LLM-Simulated Practice and Feedback Upskill Human Counselors? A Randomized Study with 90+ Novice Counselors}


\author{Ryan Louie}
\email{rylouie@cs.stanford.edu}
\affiliation{%
  \institution{Stanford University}
  \city{Stanford, CA}
  \country{United States}
}
\author{Raj Sanjay Shah}
\email{rajsanjayshah@gatech.edu}
\affiliation{%
  \institution{Georgia Institute of Technology}
  \city{Atlanta, GA}
  \country{United States}
}

\author{Ifdita Hasan Orney}
\email{ifdi1101@stanford.edu}
\affiliation{%
  \institution{Stanford University}
  \city{Stanford, CA}
  \country{United States}
}

\author{Juan Pablo Pacheco}
\email{pacheco7@stanford.edu}
\affiliation{%
  \institution{Stanford University}
  \city{Stanford, CA}
  \country{United States}
}

\author{Emma Brunskill}
\email{ebrun@cs.stanford.edu}
\affiliation{%
  \institution{Stanford University}
  \city{Stanford, CA}
  \country{United States}
}

\author{Diyi Yang}
\email{diyiy@stanford.edu}
\affiliation{%
  \institution{Stanford University}
  \city{Stanford, CA}
  \country{United States}
}


\begin{abstract}
The growing demand for accessible mental health support requires training more counselors, yet existing approaches remain resource-intensive and difficult to scale. LLMs can realistically simulate patients and generate actionable feedback for training, but their actual impact on novice counselor skill development remains unknown. We developed an LLM-simulated practice and feedback system and conducted a randomized study with 94 novice counselors, comparing practice alone versus practice with feedback. We evaluated behavioral performance, self-efficacy, and qualitative reflections. Results showed the practice-and-feedback group improved in client-centered microskills (reflections, questions), while the practice-alone group showed no improvements. For empathy, the practice-alone group declined over time and performed significantly worse than the feedback group. Qualitative interviews reinforced these findings: feedback helped participants adopt a client-centered listening approach, while practice-alone participants remained solution-oriented. These results suggest LLM-based training systems can promote effective skill development, and combining simulated practice with structured feedback is critical for meaningful improvement.

\end{abstract}






\maketitle

\input{sections/1_introduction.tex}
\input{sections/2_related_work.tex}

\input{sections/3_system.tex}
\input{sections/4_study_methods.tex}

\input{sections/5_results.tex}

\input{sections/6_discussion.tex}
\input{sections/7_conclusion.tex}


\section{Contributors}
Ryan and Raj led the technical development efforts of CARE, the LLM-simulated training system. Ryan, Emma, and Diyi worked together to design the randomized experiment. Ryan, Ifdita, Juan Pablo, and Raj conducted the 90+ online video call sessions. Ryan was responsible for the automatic behavioral assessment of counseling skills, as well as the self-efficacy calibration analysis, while Emma and Diyi advised on analysis methods and interpretations of these quantitative results. Ifdita, Juan Pablo, and Ryan conducted the thematic analysis of participants qualitative, self-reflection data. Ryan conducted the in-depth analysis of participant perceptions of receiving LLM feedback and practicing with simulated patients. All authors helped with drafting and editing the Article manuscript, references, and figures.

\begin{acks}
This research was made possible with funding support from a Stanford Impact Labs Stage 1 Award, Stanford HAI Seed Grant, Stanford Psychotherapy and Behavioral Sciences Department Innovator Award, and Stanford CURIS paid-internship program for undergraduate researchers.

The authors have a large community of researchers to thank for their help during all stages of this project. The CARE system would have not been possible without the significant code contributions from student researchers in SALT Lab, including Meijin Li, Cheng Chang, Ananjan Nandi, Alicja Chaszczewicz, and Alan Zhang. In addition, CARE's design benefited from feedback from Bruce Arnow and others from the Stanford-PAU PsyD consortium. We thank Pannisy Zhao for assistance while conducting user studies. Statistical analyses and presentation of results were much improved due to help from Robert Kraut and Akhila Kovvuri. Thank you Yanzhe Zhang and Rose Wang for detailed discussions on finetuning the LLM classifiers. Thank you to the following individuals for providing feedback on early drafts and presentations of this work, with special thanks to Yutong Zhang, Will Held, Ella Li, Michael Ryan, Dora Zhao, Caleb Ziems, Matthew Jöerke, Joy He-Yueya, Omar Shaikh, and Kapil Garg.
\end{acks}

\bibliographystyle{ACM-Reference-Format}
\bibliography{filtered_references}

\newpage
\appendix
\input{supplement/appendix}

\end{document}

%% file: sections/1_introduction.tex
\section{Introduction}
In 2023, 22.8\% of U.S. adults (approximately 58.7 million people) experienced a mental illness~\cite{SAMHSA2024}. Yet, access to effective mental health care is severely limited by shortages of qualified providers, from psychotherapists and counselors to social workers and peer supporters~\cite{larson2016supply, modi2022exploring}. While there is increasing interest in direct-to-patient AI systems with some promising results~\cite{heinz2025randomized}, we expect the demand for human-delivered mental health support to continue to far exceed supply. The limited supply of effective therapy providers is due, at least in part, to the reliance on resource-intensive methods to train helping skills~\cite{helping_skills} and evidence-based interventions~\cite{cook2017evidence, frank2020therapist}
which require access to trainers who can simulate a client interaction~\cite{kuehne2018standardized, ay2023can} and provide expert supervision~\cite{Wheeler2007-zh, watkins2014wiley}, limiting training scale~\cite{atkins2014scaling, kuhne2020standardized}.


AI systems have been increasingly applied to counselor training as a potential solution to these scaling challenges.
Recent advances in large language models (LLMs) have enabled the simulation of patients seeking mental health support~\cite{wang2024patient, louie2024roleplay}, offering rich opportunities for practice.
The use of simulated patients is not new: in medical and nursing education, human role-plays and standardized patients are routinely used, and meta-analyses show they significantly improve skill acquisition and learner confidence~\cite{shin2015effectiveness}. Mental health training has relied on a similar tradition of human role-plays to develop core helping skills.
In parallel, AI feedback systems have progressed in automatically assessing counselor behaviors such as empathy, reflections, and active listening~\cite{xiao2015rate, sharma-etal-2020-computational, flemotomos2021automated, shah2022modeling}, generating suggested responses~\cite{sharma2023human, hsu2025helping, chaszczewicz2024multi} and explanations~\cite{rudolph2024automated, chaszczewicz2024multi}.
These feedback systems target skills from client-centered approaches~\cite{raskin2005person, miller2012motivational}, empathy, reflections, questions, and active listening, which have been shown to strengthen common factors like therapeutic alliance, a powerful predictor of therapy outcomes across therapy modalities~\cite{elliott2018therapist, wampold2015important, norcross2018psychotherapy}.
However, most evaluations have largely positioned AI as a real-time co-pilot rather than a training tool~\cite{sharma2023human, hsu2025helping}, or studied pre-LLM training systems with limited practice realism and simpler, non-generative feedback mechanisms~\cite{tanana2019development}.

Recent work has developed and tested systems integrating LLM-simulated patients for training counselors, to learn case conceptualization skills for cognitive behavioral therapy (CBT)~\cite{wang2024patient} or motivational-interviewing techniques for substance-misuse scenarios~\cite{steenstra2025scaffolding}. Their studies recruit student and expert counselors to evaluate the system output quality, usability, and desirability. While prior work highlights the promise of LLM-simulations for training counselors, to our knowledge, no study has evaluated whether these LLM-simulated training systems promote novice skill improvements. A key barrier to conducting skill development studies is first establishing the LLM-system's output quality; even advanced prompt-based approaches (e.g., providing coding manuals like the Motivational Interviewing Treatment Integrity (MITI) in context~\cite{steenstra2025scaffolding}) require validation and refinement by domain-experts.

\begin{figure*}[t]
    \centering
    \includegraphics[width=\linewidth]{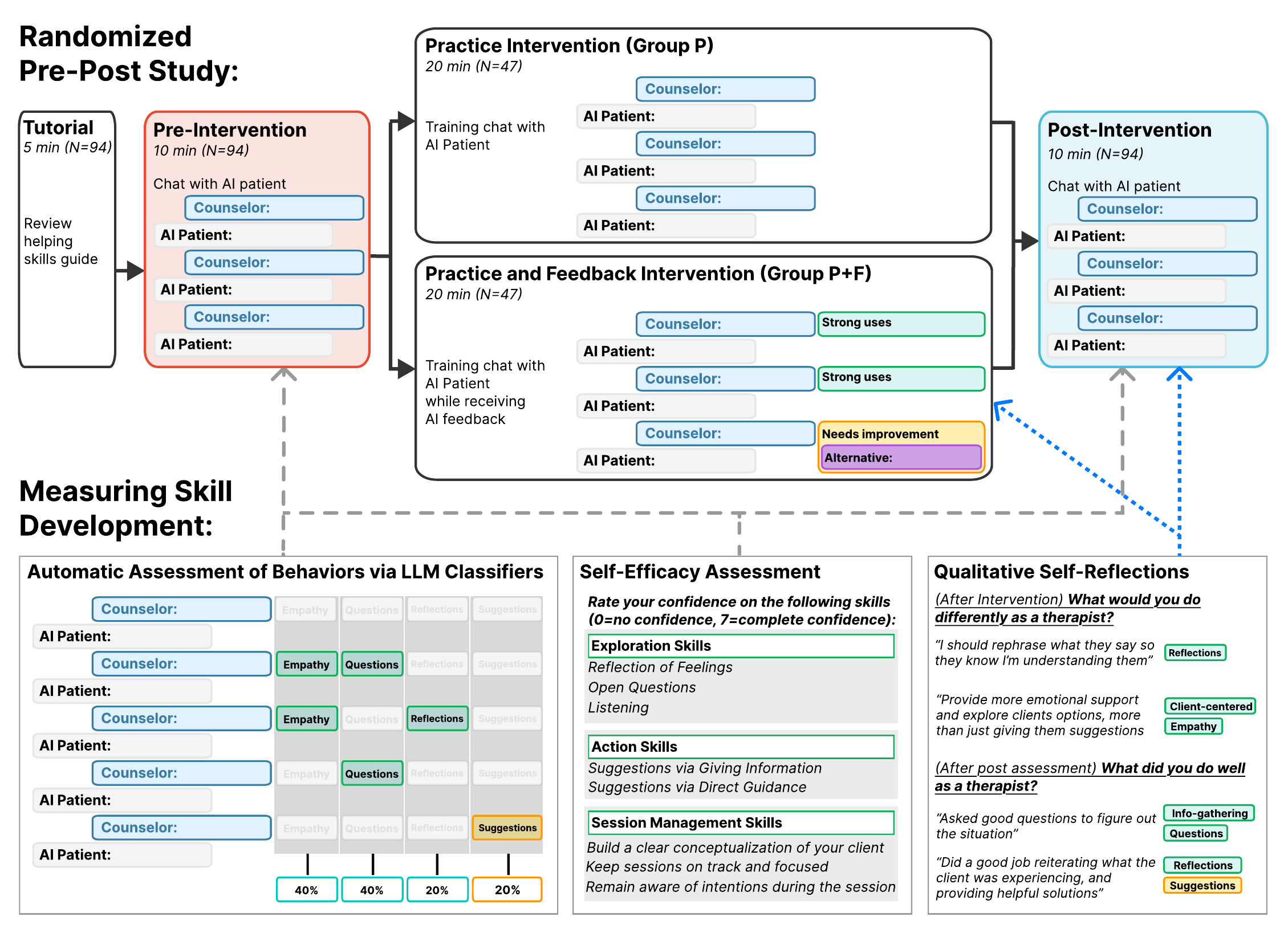}
    \caption{\small{Our experiment randomizes participants to either practice with AI Patients alone (\textbf{P}) or practice with AI patients and receive AI feedback (\textbf{P+F}). We holistically evaluate counselor skill development from three perspectives: automatic assessments of behaviors of skills used via LLM classifiers; self-efficacy and its calibration with actual performance; and qualitative self-reflections after the training intervention chat and post-intervention chat.}}
    \Description{Figure illustrating the randomized pre-post study design of CARE, an LLM-based counselor training system. Ninety-four novice counselors first completed a 5-minute tutorial and a 10-minute pre-intervention chat with an AI-simulated patient. Participants were randomized into two groups: Practice-only (P), who engaged in a 20-minute chat with an AI patient without feedback, and Practice+Feedback (P+F), who engaged in the same practice while receiving structured AI feedback, labeling strengths, weaknesses, and alternative responses across core microskills (empathy, reflections, questions, suggestions). All participants then completed a fixed 10-minute post-intervention chat with a new AI patient. Skill development was assessed across three dimensions: (1) behavioral performance, via fine-tuned LLM classifiers detecting strong or weak uses of counseling skills; (2) counseling self-efficacy, using the CASES-R survey of exploration, action, and session-management skills; and (3) qualitative self-reflections, where participants reported what they would do differently and what they did well.}
    \label{fig:studyflow}
    \Description{Figure illustrating the randomized pre-post study design of CARE, an LLM-based counselor training system. Ninety-four novice counselors first completed a 5-minute tutorial and a 10-minute pre-intervention chat with an AI-simulated patient. Participants were randomized into two groups: Practice-only (P), who engaged in a 20-minute chat with an AI patient without feedback, and Practice+Feedback (P+F), who engaged in the same practice while receiving structured AI feedback, labeling strengths, weaknesses, and alternative responses across core microskills (empathy, reflections, questions, suggestions). All participants then completed a fixed 10-minute post-intervention chat with a new AI patient. Skill development was assessed across three dimensions: (1) behavioral performance, via fine-tuned LLM classifiers detecting strong or weak uses of counseling skills; (2) counseling self-efficacy, using the CASES-R survey of exploration, action, and session-management skills; and (3) qualitative self-reflections, where participants reported what they would do differently and what they did well.}
\end{figure*}

To address this gap, we develop \system{} by combining two previous methods for realistic patient creation~\cite{louie2024roleplay} and feedback generation~\cite{chaszczewicz2024multi}, into a wholistic system for training counseling skills, and conduct a randomized experiment to investigate how two different modes of simulated training impact counselor skill development. \system enables (1) realistic practice with LLM-simulated patients, whose prompts are seeded by expert counselors to resemble challenging behaviors~\cite{louie2024roleplay}, and (2) structured feedback from a fine-tuned LLM that identifies strengths and areas for improvement across core counseling skills (e.g., empathy, reflections, questions, suggestions), while also providing explanatory rationale and alternative responses~\cite{chaszczewicz2024multi}.
The generated feedback is grounded in established counseling frameworks~\cite{helping_skills, MISC} and informed by expert-annotated examples, ensuring alignment with recognized training standards.
Uniquely, \system's LLM patients and feedback have been co-designed, iteratively improved, and rigorously validated by counseling domain-experts, ensuring that our experiment on novice skill development controls for the quality of the LLM components' outputs.

We conducted a 75-minute online lab study with 94 novice counselors to evaluate how different LLM-simulated practice modes in \system impact skill development. Participants were randomly assigned to one of two conditions: (1) \textit{Group P:} Practice with LLM-simulated patients without AI feedback, or (2) \textit{Group P+F:} Practice with LLM-simulated patients plus AI feedback (see Fig.~\ref{fig:studyflow}). We measured changes in behavioral performance (via transcript analysis), self-efficacy (via survey items), and intentions for growth (via self-reflection prompts). Our study design specifically addressed: \textit{What changes occur after practice with an AI-simulated patient alone? How do these outcomes differ when participants also receive structured AI feedback?}
Our results show the practice-and-feedback group significantly improved in their use of reflections and questions (d=0.32-0.39, p$<$0.05), and trended toward improvement in empathy (d=0.23) and suggestions (d=$-$0.28). In contrast, the practice-only group only showed significant improvements in suggestions (d=$-$0.39), but actually worsened in Empathy (d=$-$0.52, p=0.001). Between-group comparisons show a substantial advantage for the P+F group in Empathy (d=0.72, p=0.001), indicating a strong feedback effect.
In contrast, suggestions showed near-zero between-group differences (d=0.02, p=0.910) despite pre-post improvements in both conditions, suggesting that another mechanism besides feedback is driving this change.
Through qualitative analysis of participants' self-reflections, we found that the practice-and-feedback group internalized the importance of empathetic and active listening; however, practice-only participants continued to overly focus on solutions, albeit with increased information-gathering.
Our discussion details possible reasons for skill changes in the practice-alone group: novice counselors evolve therapeutic intentions from observable patient behaviors--when the AI patients consistently expressed skepticism to suggestions, counselors reduce inappropriate use of suggestions; but when AI patients show no differential response to empathetic vs. non-empathetic statements, the P group gradually de-prioritizes empathy.
Together, these results suggest that LLM-simulated training should integrate structured feedback to cultivate a client-centered, empathetic listening approach fundamental to effective counseling.

In summary, we contribute: (1) the design of \system, an LLM-based training system that integrates realistic patient simulations with structured feedback grounded in counseling frameworks; (2) evidence from a randomized evaluation with 94 novice counselors, triangulating outcomes across behavioral performance, self-efficacy, and therapeutic intentions; and (3) design implications for LLM-simulated training, showing how structured feedback prevents empathy decline and supports effective counselor development, while highlighting ongoing challenges in improving overall self-efficacy while minimizing mis-calibration with performance. 


%% file: sections/2_related_work.tex
\section{Related Work}

Our work training novice counselors via LLM-based systems is grounded in two areas of work. First, prior training approaches for clinical and communication skills have long relied on simulated patients, ranging from human role-plays to scripted virtual patients and, more recently, LLM-based simulations with automated feedback.
Second, prior HCI research evaluating human-AI systems, especially in domains like health and education, emphasizes not only AI system accuracy but also how users learn, calibrate, and reflect when interacting with AI systems.

\subsection{Training Systems for Clinical and Communication Skills}
Traditional training for helping skills—empathy, active listening, and communication—uses theory, expert demonstration, role-play, supervised practice, and experiential learning~\cite{hill2020helping, hill2002development, martin2024counseling, huerta2010experiential}. These approaches are effective but hard to scale: peers and supervisors require coordination, and trainees can pick up unhelpful habits without oversight~\cite{atkins2014scaling}.
Simulated standardized patients (trained actors) are common in health education; meta-analyses show they improve communication, knowledge transfer, and confidence~\cite{shin2015effectiveness}. Counseling training likewise uses peer role-plays and standardized scenarios to practice skills like reflective listening and empathy~\cite{kuhne2020standardized, ay2023can}, yet such exercises remain resource-intensive and limited in availability.

\paragraph{Virtual patients} Virtual patient (VP) simulations—computer or embodied agents—recreate clinical encounters to train history taking, nonverbal communication, empathy, and counseling~\cite{albright2016harnessing, schoenthaler2017simulated, rein2018evaluation}. They provide safe, repeatable practice without actors and have been applied to suicide prevention, adolescent substance-use screening, and antibiotic conversations~\cite{rein2018evaluation, burmester2019interactive, schoenthaler2017simulated}. For example, \citet{murali2022training} used a conversational agent to teach vaccination-related counseling to laypeople; commercial platforms like Skillsetter follow deliberate-practice models~\cite{skillsetter2024}. Early VP systems were often scripted and template-based, costly to develop, and typically limited to single cases, reducing realism and diversity~\cite{nguyen2003development, han2015exploiting, othlinghaus2020seriousroleplaying}.

\paragraph{Early AI-based training systems} Building on VPs, early AI-driven systems experimented with automatically analyzing communication features and providing learners with feedback. For example, EQClinic visualized audio and video signals to help trainees reflect on their nonverbal behaviors in telehealth role-plays~\cite{liu2016eqclinic}, while ConverSense detected and displayed social signals such as dominance and warmth from patient-provider interactions~\cite{bedmutha2024conversense}. These systems raised self-awareness of communication styles, but their feedback was often decontextualized and difficult to apply in practice. Moreover, they did not directly target the counseling microskills important for effective therapeutic interactions.


\paragraph{LLM-simulated patients for role-play practice} Thus, a growing body of work in NLP and HCI has used LLMs to create simulated patients as role-play partners for counselor training~\cite{wang2024patient, louie2024roleplay, stapleton2023seeing, steenstra2025scaffolding}. The goal of these systems is to provide practice environments that resemble real clinical encounters, making training more transferable and faithful to practice~\cite{alinier2022simulation}. However, achieving authentic simulations remains challenging. LLMs are highly sensitive to prompting~\cite{whyjohnnycantprompt}, and naive prompts in mental health contexts often produce unrealistic behaviors, including caricature, bias, and limited domain fidelity~\cite{cheng2023compost}.
~\citet{chen2023llmempowered} found that naively prompting GPT-3.5 to simulate a patient profile with depressive symptoms led the chatbot to describe its emotions in formal, diagnostic language, which expert clinicians noted as inauthentic. 
Recent work prompts LLMs with psychology-grounded frameworks to simulate patients (e.g., Patient-Psi for CBT case conceptualizations~\cite{wang2024patient} and multi-stage pipelines modeling cognitive factors~\cite{steenstra2025scaffolding}). Nonetheless, LLMs remain prompt-sensitive and typically need expert validation to capture realistic resistance, ambivalence, and other clinically relevant behaviors. We therefore adopt the expert-driven, iterative behavioral refinement method of \citet{louie2024roleplay}—integrating those validated principles into \system to produce more authentic simulated patients.


\paragraph{Automatic scoring and feedback for counselor transcripts}
A parallel line of work has developed automated methods to help peer counselors improve their skills. Scoring-based systems (e.g., ratio of questions to reflections in a transcript) provide metrics, but these approaches offer limited, actionable guidance on how to improve. By contrast, suggestion-based systems generate or rewrite candidate responses to model more effective behaviors. Research in clinical NLP has produced numerous models for classifying and scoring counseling transcripts~\cite{ tanana2016comparison, perez2019makes, schwalbe2014sustaining, huang2018modeling, fang2023makes, min-etal-2022-pair}. Many focus on a single microskill, such as \emph{reflections}, providing numeric feedback on usage frequency~\cite{can2012case, perez2017predicting, shen2020counseling, min-etal-2022-pair}. Others examine skill distributions more broadly and their relationship to conversational success~\cite{wang2023metrics}. While valuable for large-scale analysis, these approaches rarely translate into actionable feedback for scaffolding trainee learning.
To make feedback more interactive, researchers have explored real-time rewriting and suggestion systems. For example, \citet{saha2022towards} and \citet{sharma2021towards} proposed response rewriting methods to enhance empathy. With the goal of increasing interactiveness, ~\citet{sharma2023human} proposed HAILEY, a tool that modifies peer supporters’ responses, while ~\citet{hsu2025helping} generated strategy-aligned suggestions during live conversations. Although promising, studies show that just-in-time suggestions can distract learners and foster overreliance, sometimes leading to negative learning effects when AI support is withdrawn~\cite{abbas2024harmful, bastani2024generative, kazemitabaar2024codeaid}.

While this previous work developed NLP models for specific counseling tasks, the ability to use LLMs as zero-shot or few-shot reasoners has enabled further research in this area.
Nonetheless, naively prompting LLMs in a mental health context can lead to generated outputs that are characteristic of low-quality therapy~\cite{chiu2024computational}. Therefore, a training system that uses LLMs to generate feedback for counselors needs to take measures to ensure the outputs are faithful and robust, lest it teach or promote bad practices \cite{moran2025artificial}.
\citet{chaszczewicz2024multi} co-designed a feedback dataset with therapy supervisors and fine-tuned an open-weight LLM to produce explanatory, actionable feedback. \system adopts this expert-validated, fine-tuned model. While other training systems have employed prompting-based approaches with coding manuals and few-shot examples~\cite{steenstra2025scaffolding}, our approach differs by integrating expert feedback directly into model weights through fine-tuning on expert-annotated examples, rather than relying on in-context learning alone.

\subsection{Evaluating Human-AI Systems}

Evaluating human-AI systems requires more than assessing model accuracy or output quality ~\cite{blodgett2024human}. In HCI and education research, effectiveness is judged by its impact on learners: how people acquire skills, calibrate their understanding, and integrate feedback into practice. This framing is important in counseling training, where evaluation concerns not only usability but also the development of interpersonal behaviors in sensitive, high-stakes domains.

Recent work highlights the limitations of traditional benchmarks, which often fail to capture generative model capabilities~\cite{mcintosh2025inadequacies}. This calls for dynamic and human-centered evaluations \cite{liao2023rethinking, elangovan2024considers, khullar2025nurturing}, that move beyond static model metrics and consider human outcomes, and are relevant when assessing interactive training systems. Thus, when we evaluate human-AI systems, additional challenges arise. Researchers must account for both the technical performance and also user impact~\cite{weidinger2023sociotechnical}. While guidelines exist for designing human-AI systems~\cite{amershi2019guidelines,wright2020comparative}, less work addresses how they should be evaluated. Some frameworks capture process and user preferences in human-LLM interaction~\cite{lee2022evaluating}, others focus on safety~\cite{weidinger2023sociotechnical} or domain-specific contexts~\cite{lee2024design}. Tools such as SPHERE propose multi-dimensional evaluation cards to structure study design and improve transparency, but consensus on evaluation practices remains limited \cite{ma2025sphere}.

In counseling contexts, these gaps surface in three ways. First, evaluation must triangulate across behavioral outcomes, self-efficacy, and learning, aligning with evidence-based psychotherapy work in deliberate practice~\cite{ericsson2006influence, schon2017reflective}. Second, calibration is critical: learners often misjudge their own performance~\cite{kruger1999unskilled,eva2005self}, and AI feedback may inflate confidence without improving skills~\cite{macnamara2024does}. Third, user perceptions of realism, trust, and workload shape adoption: relational agent studies show that authenticity fosters engagement~\cite{mallik2023proactive}, while trust research highlights risks of distraction and overreliance~\cite{glikson2020human,buccinca2021trust}. Finally, in sensitive domains, evaluation must weigh ethical and pedagogical guardrails: ensuring feedback preserves learner agency and avoids harmful or misleading guidance.

Taken together, evaluating human-AI systems requires a multi-dimensional perspective that integrates skill outcomes, calibration, perceptions, and responsible design. Yet, few studies have examined how LLM-based training systems affect novice counselors across these dimensions. Our work contributes by combining objective performance measures, self-efficacy surveys, and qualitative reflections to provide a holistic evaluation of LLM-driven counseling training. 

%% file: sections/3_system.tex
\section{\system Training System}
\begin{figure*}[!h]
    \centering
    \begin{subfigure}{\textwidth}
        \includegraphics[width=0.9\linewidth]{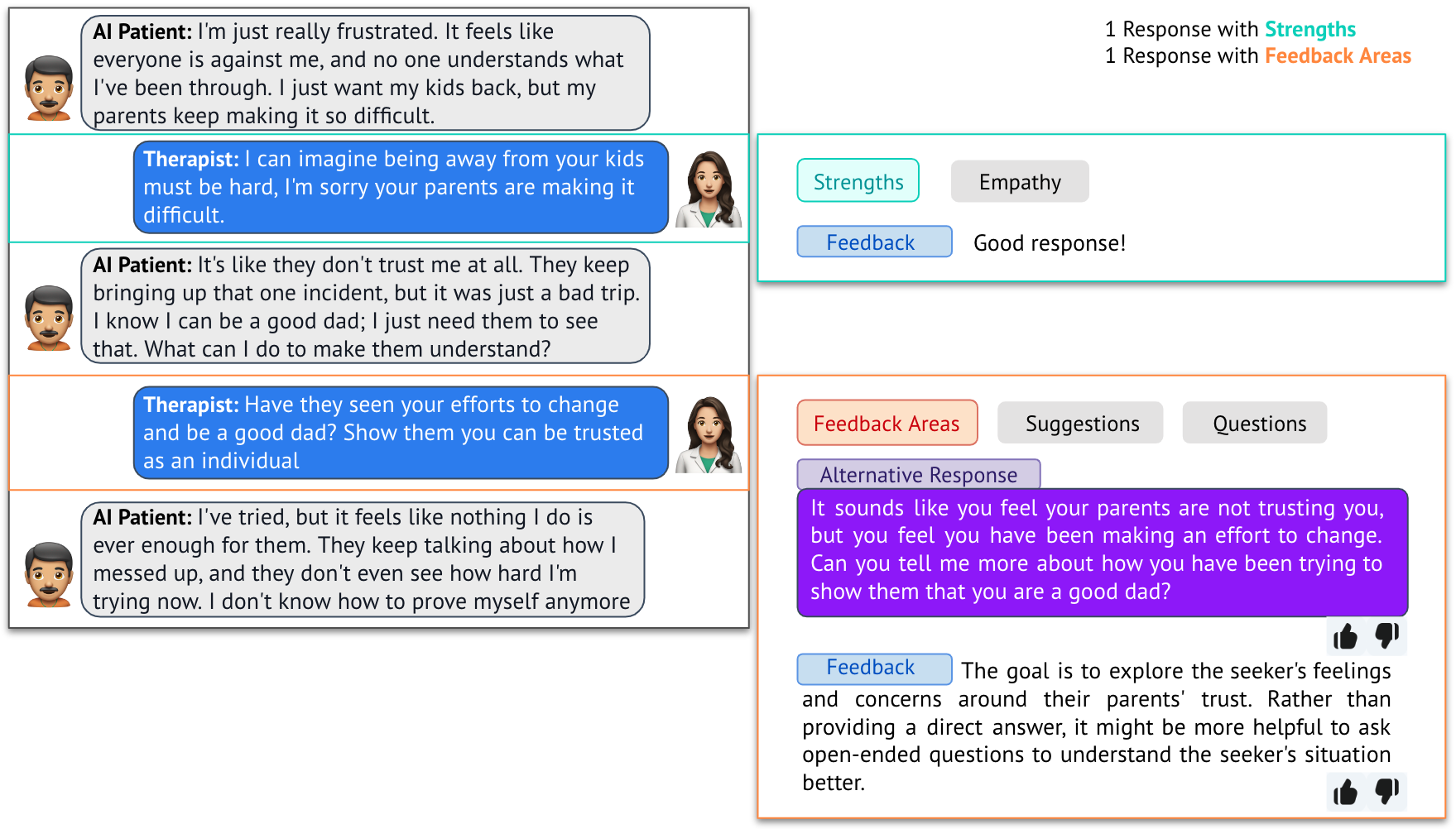}
    \end{subfigure}
    \begin{subfigure}{\textwidth}
        \includegraphics[width=0.9\linewidth]{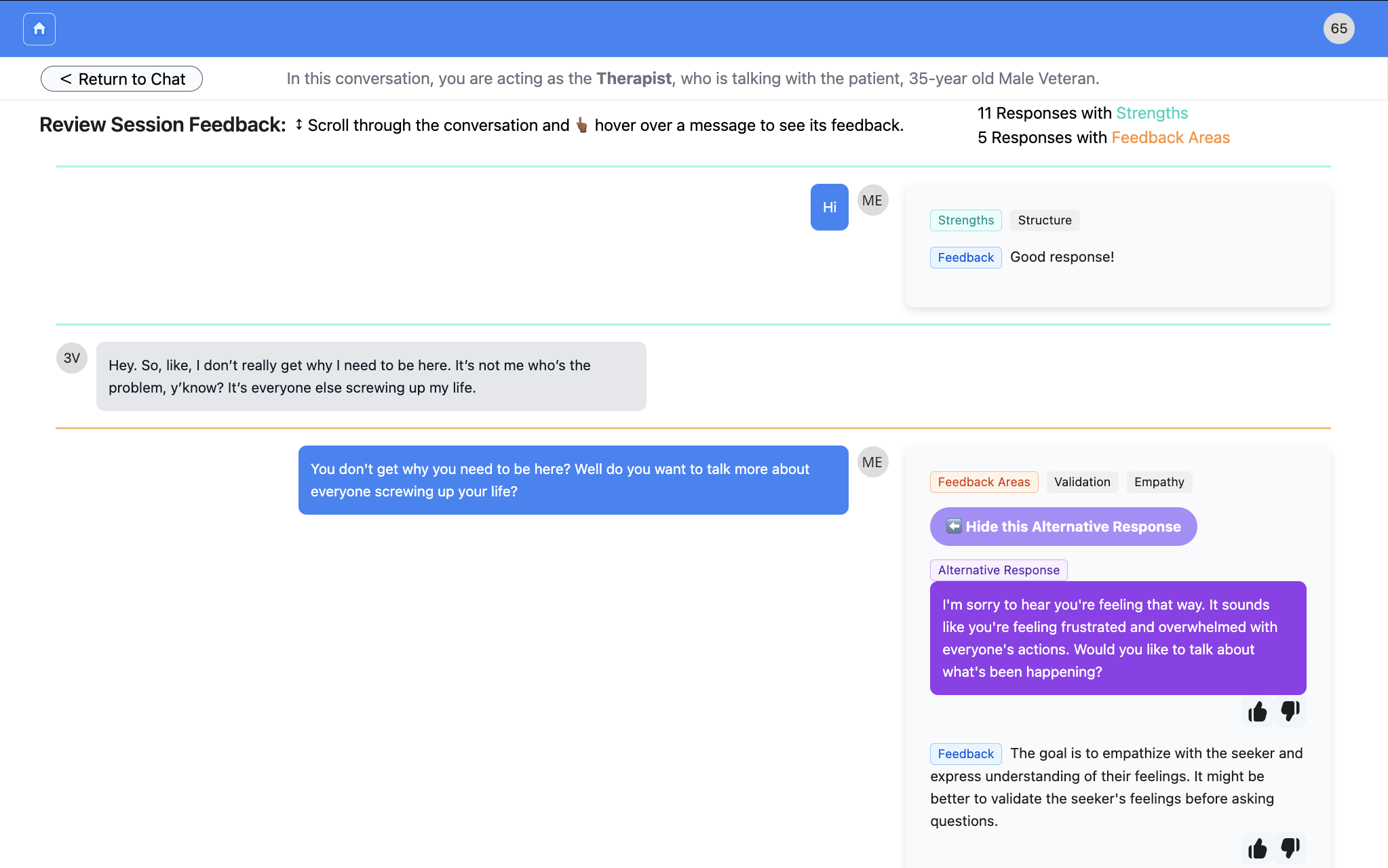}
    \end{subfigure}
    \caption{\system's practice and feedback model visualized in a web screenshot. In \system, counselors practice with an LLM-simulated patient and receive feedback on each of their responses. The feedback model labels whether a response has \textcolor{teal}{\textbf{strengths}} or constructive \textcolor{orange}{\textbf{feedback areas}}.
    Responses with constructive \textcolor{blue}{\textbf{feedback}} explain what the goal should be at this point in the conversation; what a helper could improve to better align with this goal; and how they could respond differently via an \textcolor{violet}{\textbf{alternative response}}.}
    \label{fig:care-screenshot-reviewfeedback}
    \Description{The figure shows a screenshot of CARE’s practice and feedback interface. On the left, a text-based chat unfolds between an AI patient (depicted with a client avatar) and a counselor (depicted with a clinician avatar). The AI patient expresses frustration about his parents' lack of trust and his struggles to prove he is a good father. The counselor responds twice: first with an empathetic statement that acknowledges the difficulty, and second with a suggestion-oriented prompt about showing efforts to change. On the right, CARE’s feedback system evaluates these responses. The first counselor response is labeled a Strength, tagged as “Empathy,” with positive feedback saying “Good response!” The second counselor's response is flagged as a Feedback Area, tagged under “Suggestions” and “Questions.” The system provides an Alternative Response that rephrases the counselor’s statement to be more client-centered and exploratory, asking about the patient’s efforts rather than giving advice. Additional Feedback text explains that the goal at this point is to explore the seeker’s feelings and concerns rather than provide direct solutions, encouraging open-ended questioning.}
\end{figure*}

We developed \system as a web platform for novice counselors to train in text-based counseling skills enabled by LLMs. The system integrates two core components: (1) LLM-simulated patients that provide realistic, text-based practice conversations, and (2) LLM-generated feedback that evaluates counselor responses against established skill frameworks and suggests improvements. Together, these features enable scalable, authentic training experiences that complement traditional, resource-intensive approaches such as role-play and supervision.

\system builds on top of successes from previous research in co-designing with mental health experts to improve the realism of LLM-simulated patients~\cite{chen2023llmempowered, lin2024imbue, wang2024patient, louie2024roleplay}, using fine-tuned domain-specific LLMs trained on therapeutic knowledge capable of generating feedback and alternative responses for text-based peer counseling conversations~\cite{sharma2021towards, chaszczewicz2024multi, shen-etal-2022-knowledge, min-etal-2022-pair, min-etal-2023-verve}. Importantly, \system was designed not only to identify whether a skill is used but also how well it is used, distinguishing, for example, between a reflection that captures a client’s core concern and one that misses the emotional nuance.
\system allows novice counselors to develop their counseling skills in a text-based format by practicing with AI-simulated patients and receiving feedback on their responses (see Fig.~\ref{fig:care-screenshot-reviewfeedback}).

\begin{quote}
\textit{Consider Aki, a novice peer counselor who wants to use \system to experience hands-on training using counseling skills that they have recently read about. In \system, Aki can practice with an AI patient of their choice from a library of patient scenarios. Aki initiates a practice chat with an AI patient, a 35-year-old male veteran who is seeking to reconnect with his children but is facing legal barriers and parental gatekeeping.}
\end{quote}

Each practice scenario provides limited background information about the AI patient and their presenting problem (e.g., "Young adult with family issues: low mood and self-esteem"). This intentional limitation requires counselors to simultaneously learn more about the patient's situation while demonstrating empathy and support. Patients are designed to exhibit realistic challenges, including resistance, ambivalence, or vague disclosures, drawing on behavioral principles elicited from expert counselors~\cite{louie2024roleplay}.

\begin{quote}
\textit{Aki starts the conversation by greeting the AI patient. The AI patient expresses frustration about feeling that everyone is against them, hoping to find ways to reunite with their kids and overcome the challenges posed by their judgmental parents. Aki composes a reply, and the conversation continues in this turn-by-turn manner.}
\end{quote}

\begin{quote}
\textit{After completing the conversation, Aki reviews AI-generated feedback. \system highlights strengths such as asking open-ended questions, but also flags missed opportunities for empathy, offering alternative phrasings, and a rationale for why they may better support the client.}
\end{quote}

Unlike real-time AI "co-pilot" systems, which may proactively suggest responses, \system provides feedback only after the user has sent their response in the simulated dialogue. A user can view feedback on their therapeutic responses at any point. This offers flexibility to review feedback intermittently throughout the practice or comprehensively after completion. This design choice mirrors human supervision: it preserves the learner’s agency during the conversation while supporting reflection afterward. Feedback targets core microskills, empathy, reflections, questions, validation, and suggestions, based on established counseling frameworks~\cite{helping_skills, MISC}.

\subsection{Implementation details for \system Training Platform}

\system was built as a web application with a Python Flask back-end and React JavaScript front-end, accessible through any standard browser. The two core components of \system are described below.

\textbf{LLM-simulated patients.} \system implements an existing method for simulated patients that co-designed patient prompts with expert-counselors, each specifying (a) a challenging patient scenario that they had encountered in past clinical and text-based counseling settings, including demographic background presenting issues or symptoms, etc. and (b) Constitutional AI principles elicited from expert counselors for defining authentic patient behaviors~\cite{louie2024roleplay}.
This simulation method uses the OpenAI \texttt{GPT-4o} API to role-play patient scenarios and behaviors due to its strong ability to maintain role consistency and instruction-follow expert-defined principles. These principles instructed the simulated patients to display realistic challenges such as resistance to suggestions (\texttt{"Respond to encouraging words with hesitation, doubting their significance"}), low awareness (\texttt{"Don't be so self-aware or good at recognizing your own problems"}), or minimal disclosure (\texttt{"Use more colloquial language and express reluctance
to open up"}). Simulated patients were designed and validated for 10-20 minute text-based counseling conversations, allowing participants to both explore the presenting problem and practice multiple client-centered microskills. \system implements the top-rated patients from~\citet{louie2024roleplay}'s study which were judged by third-party counselors as being the most ready to be used as a training partner ($\geq6$ average score on a 7-point Likert-scale). All prompt templates are available in Supplementary Materials~\ref{sec:llm-patient-prompts}.

The three AI patients used in our experiment were created by experienced counselors from the USA with extensive backgrounds in mental health support~\cite{louie2024roleplay}. Each patient was deliberately varied to expose trainees to diverse yet comparably challenging scenarios:
\begin{itemize}
\item \textbf{Patient 1 (35-year-old American male experiencing holiday loneliness):} Created by an experienced peer counselor from an online counseling platform. Holiday loneliness is common there and often stems from family estrangement. In this case, the patient's isolation intensifies during winter holidays when family gatherings occur.
\item \textbf{Patient 2 (35-year-old male veteran in court-mandated therapy seeking to reconnect with children):} Created by a Licensed Marriage and Family Therapist (LMFT), a white woman aged 30-40, based on her clinical experience with veterans. This scenario reflects a patient frustrated by limited access to his children due to parental and legal issues stemming from substance abuse.
\item \textbf{Patient 3 (Young adult with family issues, low mood, and self-esteem concerns):} Created by a US-basd clinical psychology doctoral student. This adolescent patient faced self-esteem issues stemming from family dynamics, where her parents favored a sibling, leading to symptoms of anhedonia and depression, with a diminished ability to enjoy previously pleasurable activities.
\end{itemize}

\textbf{LLM-generated feedback.} \system{} integrates an existing method for generating counseling feedback that fine-tunes and self-improves the Llama-2 13B parameter model using an expert-annotated feedback dataset of peer counseling transcripts ~\cite{chaszczewicz2024multi}. We selected this existing method because it provides faithful generation of counseling feedback grounded in the content and style of feedback given by psychotherapy supervisors. An additional key benefit, as argued by~\citet{chaszczewicz2024multi}, is that a fine-tuned open-weight model can operate on therapy data in a controlled, private environment rather than relying on external API services. This model generates feedback at multi-levels: (1) classify the trainee response against eight microskills (empathy, reflections, questions, validation, suggestions, session management, professionalism, and self-disclosure), (2) assess quality by highlighting strengths and areas needing improvement, and (3) generate alternative responses and explanatory rationales, enabling trainees to compare their choices against more client-centered approaches. This post-practice feedback design mirrors human supervision: it preserves agency during the conversation while supporting reflection and skill refinement afterward.



%% file: sections/4_study_methods.tex
\section{Randomized Pre-Post Study}

The core goal of \system is to upskill novice counselors through LLM-simulated training.
In a randomized experiment, we investigated how \system's core components--\textbf{practicing} with LLM-simulated patients and receiving AI-generated \textbf{feedback} on their responses--are important for participants' skill development. We conceptualize skill development holistically, encompassing three complementary dimensions: (1) \textit{behavioral performance}, where a trainee is judged on their appropriate use of counseling skills in a representative scenario or conversation; (2) \textit{counseling self-efficacy}, defined as a trainee's self-assessments of their own abilities; and (3) \textit{therapeutic intentions}, or the goals that participants form in-session, which should be adherent with evidence-based procedures.

The experiment evaluated how skill development changed over time when participants were assigned to two variants of LLM-based counseling training: practicing with LLM-simulated patients alone (Group P); or practicing with simulated patients while also receiving LLM-generated feedback (Group P+F). In contrast to other work on simulation in clinical education~\cite{shin2015effectiveness}, our primary interest was in understanding what aspects of LLM-simulated training could promote skill development, as part of a broader research agenda to iteratively design and improve AI-based counseor training. Given this, we did not include a "no AI" control group.

Beyond the skill development outcomes, we also conduct a mixed-methods investigation of participants' experience of \system's LLM-based components and their perceived value of such training experience, since user perceptions shape adoption in training contexts.
Together, our experiment sought to answer six research questions covering skill development and training experience with \system from quantitative and qualitative perspectives.

\begin{itemize}
\item RQ1: How does \system's LLM-simulated practice and feedback affect participants' \textit{behavioral performance}?
\item RQ2: How does \system's LLM-simulated practice and feedback affect participants' \textit{self-efficacy}?
\item RQ3: How does \system's LLM-simulated practice and feedback affect novice counselors' \textit{therapeutic intentions}?

\item RQ4: What are participants' \textit{quantitative experience} of \system's LLM-simulated practice and feedback?
\item RQ5: What are participants' \textit{qualitative experience receiving feedback on their responses} from \system?
\item RQ6: What are participants' \textit{qualitative experience practicing} with \system's LLM patients?
\end{itemize}

\subsection{Participants}



We recruited $N=94$ novice counselors on the Prolific platform using specific filtering criteria to select US and UK participants with some interest in the field but limited access to formal training. Eligible participants were required to have (1) an educational background in psychology, counseling, social work, or nursing, with educational attainment limited to those who had completed at most a bachelor's degree or were currently pursuing a master's degree, and (2) less than one year of counseling-related experience (e.g., peer support or crisis counseling volunteering). Prolific participants were paid \$15/hour.
We conducted sessions with 108 participants from Prolific. The first 14 were part of a pilot that refined our recruiting criteria and protocol (e.g., excluding counselors with graduate degrees). Our analyses use the final 94 participants, though we reference participants by their original identifiers.
For the final participant pool, 68\% were located in the United States and 32\% in the United Kingdom. The sample was predominantly female (68\%), with 31\% male participants and 1\% preferring not to disclose gender. The median age was 29 years (IQR: 23-39). Regarding ethnicity, 49.5\% of participants identified as White, 16.2\% as Black, 15.3\% as Multiracial, 13.5\% as Asian, and 5.4\% as Other. Participants' primary fields of study included psychology (66\%), social work (24\%), nursing (16\%), and counseling (10\%), with participants able to select multiple areas.  In terms of educational attainment, 22.4\% had no formal education in these fields, 50.6\% were currently pursuing undergraduate degrees, 12.9\% had completed only bachelor's degrees in relevant fields, and 14.1\% were pursuing master's degrees.

To protect participant identities, our IRB-approved protocol instructed participants to use their Prolific email address, turn off cameras, and change their Zoom display name to their ProlificID. During recruitment and consent, we warned participants of the potentially stressful nature of simulated patient situations and ensured scenarios avoided especially sensitive topics such as suicidal ideation.

\subsection{Power Analysis}
To determine the appropriate sample size for our randomized pre-post study, we conducted a power analysis targeting a medium effect size with adequate statistical power. Our analysis was based on a repeated-measures design comparing pre-intervention and post-intervention outcomes between two groups (practice-only vs. practice-with-feedback). We selected an effect size of $d=0.4$ as our target, representing a conservative estimate for behavioral skill improvements. This choice was informed by previous research on social skills training interventions, where studies examining changes in behavioral performance have reported medium to medium-high effects ranging from $d=0.5$ to $d=0.6$~\cite{lin2024imbue}. 
Using power analysis calculations for two-sample, repeated-measures designs with $\alpha=0.05$ and $\beta=0.8$ (80\% power) in the R statistical analysis software, our calculations indicated that $N=94$ participants would provide sufficient statistical power to detect our target effect size. 

\subsection{Study Setup}
The study flow is illustrated in Figure~\ref{fig:studyflow}. The 75-minute study was conducted over Zoom. Participants first read a 5-minute tutorial refreshing foundational counseling skills, then completed a timed 10-minute pre-intervention chat with the first AI patient. For the 20-minute main intervention, we randomized participants into two groups: (1) \textit{Group P:} Practice with an LLM-simulated patient without AI feedback, or (2) \textit{Group P+F:} Practice with an LLM-simulated patient with AI feedback. Group P+F participants could review AI feedback on their responses at any time. The experimenter provided verbal reminders to check feedback at 5 minutes and to review remaining feedback at 15 minutes. Participants then completed a 10-minute chat with the third AI patient. Surveys were administered after each chat period.
Upon completing the post-intervention chat and self-efficacy assessment, participants shared their experience with the \system training tool via survey and semi-structured interview. Group P participants received 5 additional minutes to interact with AI feedback on their post-intervention chat before sharing perceptions. Since this occurred after the skill acquisition experiment, it does not interfere with training effectiveness results (RQ1-3) but allows us to ask all 94 participants their perceptions of both AI patients and AI feedback in \system (RQ4).

\subsection{Measures}
To understand whether simulated practice alone (P) and practice with feedback (P+F) can upskill novice counselors, we integrate evidence from three sources of data: automatic assessments of behavioral performance (RQ1), participants' assessments of their self-efficacy (RQ2), and qualitative self-reflections about their therapeutic intentions (RQ3).
Following the post-intervention, we conducted a final survey and semi-structured interview with participants to understand their perceptions of the \system system and its features (RQ4).

\subsubsection{\textbf{RQ1. Automatic Assessment of Behavioral Performance}}
We assess whether counselors employ higher-quality counseling behaviors in transcripts by leveraging NLP methods.
This automatic assessment is motivated by the need to quantify changes in counseling skill use at scale across multiple participant sessions.
Our automatic assessment approach requires (1) fine-tuning and validating LLM-based classifiers to identify skill behaviors, and (2) selecting a final set of classifiers based on statistical-testing considerations, performance metrics, and theoretical priority. In the following paragraphs, we explain both of these steps in more detail. Ultimately, we assessed behaviors of skills used for the exploration stage (strong uses in empathy, reflections, questions) and action stage (suggestions needing improvement) of Hill's Helping Skills framework~\cite{helping_skills}; see Table~\ref{tab:skill_descriptions} for definitions.

\paragraph{Fine-tuning and Validating LLM-based Classifiers.}
We developed LLM-based binary classifiers to label skill use within transcripts. For example, one classifier determines which utterances showed strong use of Questions. To finetune and evaluate these classifiers, we transformed a previously published expert-annotated feedback dataset~\cite{chaszczewicz2024multi} into 16-class binary classification format (8 skills × 2 categories: strong uses and areas needing improvement).\footnote{The binary classification dataset for FeedbackESConv can be accessed at \url{https://huggingface.co/datasets/youralien/feedback_qesconv_16wayclassification}}
Additionally, we used a subset of transcripts from this study, annotated by three counseling domain-experts: a \textit{practicing clinical psychologist}, \textit{licensed marriage family therapist}, and \textit{former director and supervisor of a crisis agency}. Each expert annotated 10 participants' transcripts (5 from each group), totaling 370 counselor utterances. After an initial pass, we showed experts each other's annotations for disagreement points and had them re-annotate with rationales. Table~\ref{tab:skill-agreement-performance} shows pairwise agreement results averaged across all pairs. While we initially explored metrics like Cohen's kappa, severe class imbalance made them less relevant. The {\em \system expert-annotated sample} (10\% of participants' transcripts) consists of majority-vote labels across the three experts.\footnote{This expert-annotated data sample can be found at \url{https://huggingface.co/datasets/youralien/CARE_10percent_16wayclassification}}

We finetuned RoBERTa-large binary classifiers using FeedbackQESConv, a dataset of transcripts from emotional support conversations between peer counselors on a crowdsourcing platform annotated with multi-level counseling feedback~\cite{chaszczewicz2024multi}, allocating 95\% of this data for training. For hyperparameter tuning, we used a validation set comprising 5\% of the FeedbackQESConv dataset (n=409) combined with our \system expert-annotated sample (n=370). The performance of our LLM-based classifier candidates varied by skill, as shown in Table~\ref{tab:skill-agreement-performance}, motivating us to down-select a final set of classifiers for our planned analyses.

\begin{table*}[t]
\centering
\begin{tabular}{l|c|cc|c|cc}
\toprule
\multirow{3}{*}{\textbf{Skill}} & \multicolumn{3}{c|}{\textbf{Strengths}} & \multicolumn{3}{c}{\textbf{Areas to Improve}} \\
\cmidrule{2-7}
 & Annotator & \multicolumn{2}{c|}{Classifier} & Annotator & \multicolumn{2}{c}{Classifier} \\
 & Agreement & \multicolumn{2}{c|}{Performance} & Agreement & \multicolumn{2}{c}{Performance} \\
 & \% & acc. & f1 & \% & acc. & f1 \\
\midrule
Empathy & 0.793 & 0.813 & \textbf{0.741} & 0.809 & 0.859 & 0.389 \\
Reflections & 0.863 & 0.900 & \textbf{0.562} & 0.944 & 0.903 & 0.312 \\
Questions & 0.732 & 0.784 & \textbf{0.775} & 0.852 & 0.842 & 0.394 \\
Suggestions & 0.919 & 0.955 & 0.507 & 0.946 & 0.941 & \textbf{0.681} \\
Validation & 0.726 & 0.852 & 0.556 & 0.919 & 0.893 & 0.265 \\
Self-disclosure & 0.982 & 0.920 & 0.326 & 0.969 & 0.986 & 0.849 \\
Session Management & 0.968 & -- & -- & 0.941 & -- & -- \\
Professionalism & 0.905 & -- & -- & 0.969 & -- & -- \\
\bottomrule
\end{tabular}
\caption{Annotator agreement columns show pairwise agreement averaged across 3 domain-experts for the \system expert-annotated sample (n=370). Classifier performance columns show performance of the best RoBERTa-large classification models after hyperparameter tuning on our validation dataset, \system expert-annotated sample (n=370) + FeedbackQESConv 5\% sample (n=409). Session Management and Professionalism were excluded from finetuning due to infrequent occurrence.}
\label{tab:skill-agreement-performance}
\end{table*}

\paragraph{Down-selecting a Final Set of Classifiers}
From the initial set of 16 binary classifiers, we narrowed our focus to four key classifiers: strong uses of Empathy, Reflections, and Questions, and inappropriate uses of Suggestions. This selection was guided by selecting (1) the \textit{highest performing} classifiers based on F1 scores (2) fewer classifiers for \textit{statistical concerns} since our analyses would control for a false discovery rate based on number of skill hypotheses tested; and (3) \textit{theoretical relevance} of skills most frequently emphasized in client-centered counseling textbooks and used during training with \system. The final four skill classifiers have F1 performance scores between 0.56 and 0.77. Detailed selection criteria and rationale are provided in Appendix~\ref{app:classifier-selection}.

\subsubsection{\textbf{RQ2. Counseling Self-Efficacy}}
To measure counselor self-efficacy, we employed the Counselor Activity Self-Efficacy Scale (CASES)~\cite{lent2003development}, specifically utilizing a revised subset of items targeting basic counseling skills (CASES-R)~\cite{hahn2021assessment, hunsmann2024basic}. The CASES-R established a three-factor structure to assess counselors' confidence in performing key therapeutic functions: \textit{Exploration and Insight Skills}, \textit{Action Skills}, and \textit{Session Management Skills}.

Participants completed the CASES-R immediately following both pre-intervention and post-intervention AI patient interactions. All items were administered using an 8-point Likert scale (0 = no confidence, 7 = complete confidence). During factor analysis, we discovered that among the five original Exploration and Insight Skills, self-disclosure did not load on the same factor as the other items. Consequently, we consolidated the Exploration and Insight Skills dimension to include only four Exploration Skills: Reflections of Feelings, Restatements, Open Questions, and Listening.

The final instrument comprised 12 items across three factors: (1) \textit{exploration skills} (e.g., restatements, reflecting feelings, open questions, listening); (2) \textit{action skills} (e.g., providing suggestions, knowing which actions to take); and (3) \textit{session management skills} (e.g., keeping sessions on track). To assess the internal consistency of each factor, we conducted reliability analysis using Cronbach's $\alpha$, which measures how closely related a set of items are as a group~\cite{tavakol2011making}. The analysis demonstrated good to excellent internal consistency across all factors, with Cronbach's $\alpha$ values of 0.784, 0.803, and 0.905 for exploration skills, action skills, and session management skills, respectively.

\begin{table*}
    \footnotesize
    \centering
    \begin{tabular}{|p{0.5cm}|p{0.5cm}|p{2.75cm}|p{7.5cm}|} \hline
        \multicolumn{2}{|c|}{\textbf{Stages}} & \textbf{Skill Category} & \textbf{Description} \\ \hline
        \cellcolor{red!60} & \multirow{3}{*}{} & Questions & Questions seek information from the client and can be open (inviting elaboration) or closed (requesting specific answers). They include both direct questions and indirect prompts (e.g., "Tell me about…"). \\ \hline
        \cellcolor{red!60} & & Reflections & Reflections capture and return to clients something they have communicated, either explicitly or implicitly. They typically mirror back content from the client's preceding statement, but can also reference earlier parts of the conversation. \\ \hline
        \vspace{0.1cm}
        \cellcolor{red!60} & & Empathy & Empathy can be shown through emotional warmth, interpretation of the client's experience (e.g., paraphrasing, making conjectures, or sharing relatable experiences), or exploration of the client's feelings and perspectives. \\ \hline
        & \cellcolor{green!60} & Suggestions & Suggestions offer possible actions, perspectives, or solutions in a respectful and autonomy-supportive manner. They may involve information-sharing or proposing alternative viewpoints. \\ \hline
        \multicolumn{2}{|c|}{\cellcolor{gray!60}} & Session Management & Session management includes organizing the session, transitioning between topics, and summarizing key points. It provides structure and helps maintain therapeutic focus. \\ \hline
    \end{tabular}
    \caption{\small{
    Overview of our analysis of skill development, grounded in Hill's Helping Skills model~\cite{helping_skills}. We select a skill subset relevant for beginning counselors at the undergraduate and first-year graduate level~\cite{hunsmann2024basic}. These include microskills during the \colorbox{red!60}{\textit{exploration}} and \colorbox{green!60}{\textit{action}} stages; and \colorbox{gray!60}{macro} skills that are applicable throughout the session. Hill's \textit{insight} stage, of which self-disclosure was the only relevant skill for basic counseling, was excluded from our primary analyses due to its infrequent occurrence in our data.}}
    \label{tab:skill_descriptions}
\end{table*}

\subsubsection{\textbf{RQ3. Qualitative Self-Reflections on Therapeutic Intentions}}

LLM-simulated training provides opportunities for experiential learning~\cite{huerta2010experiential} whereby reflection on action~\cite{schon2017reflective} can support counselors in refining their therapeutic intentions and strategies.
To study this impact on participants' intentions, we collected qualitative self-reflections from two time points: immediately after the training intervention chat, where participants responded to \textit{"What would you do differently as a therapist?"} and after the post-intervention chat, where they reflected on \textit{"What did you do well as a therapist?"}. We examined how initial intentions translated into reported strengths across the P+F and P groups.


\subsubsection{\textbf{RQ4. Quantitative Experience of \system's LLM Feedback and Simulated Practice}}
Three survey questions measured participants' perceptions of \system's AI feedback system on a 5-point Likert scale.
\textbf{Helpfulness.} Participants rated \textit{"To what extent do you find the AI feedback to be constructive and helpful?"}. \textbf{Comfort.} Participants rated \textit{"To what extent do you agree with the following statement: 'I am comfortable receiving AI feedback'"}. \textbf{Readiness.} Participants rated \textit{"The AI feedback system is ready to be used by counselors-in-training."} Participants in the P+F group answered these questions after the intervention chat. Participants in the P group also received feedback--only after the pre-post experimental measures were completed--and subsequently answered these three questions about AI feedback.

We measured participants' perceptions of each of the AI patients after each chat (pre-intervention, practice intervention, post-intervention) with several 7-point Likert scale items. \textbf{Authenticity.} Participants rated \textit{"The AI patient was authentic in its role."} Four questions from the NASA-TLX workload scale were given after each simulated practice: \textbf{Mental Demand:} "How mentally demanding was giving counseling support to this patient?"; \textbf{Temporal Demand:} \textit{"How hurried or rushed did you feel giving counseling support to this patient?"}; \textbf{Effort:} \textit{"How hard did you have to work to accomplish your level of performance"}; \textbf{Frustration:} \textit{"How discouraged or stressed were you while giving counseling support to this patient?"}.

\subsubsection{\textbf{RQ5 and RQ6. Qualitative Experience of using \system's LLM Feedback and Simulated Practice}}
Participant interview data was collected at different time-points throughout the 75-minute session. To specifically understand the experience receiving feedback from \system (RQ5), participants were asked to elaborate in more detail their answers after filling out the three survey questions about \system's feedback; re-review the AI feedback page and think-aloud about their agreements or disagreements with any of the feedback; and explain whether the feedback had any bearing on their self-reflections about what they did well or wanted to do differently as a counselor. To specifically understand the experience practicing with \system's simulated patients (RQ6), participants were given the chance to explain in more detail their answers to the quantitative survey items about the simulated patients. Finally, a semi-structured exit-interview was conducted for all participants which was framed around the following questions: \textit{"What do you like about this training tool for helping skills?"}  \textit{"What do you wish was different about the training tool?"} and \textit{"What suggestions do you have for improving any part of the training tool?"}.

\subsection{Analyses}

\subsubsection{\textbf{RQ1. Effects on Behavioral Performance}}
Our analysis of behavioral performance consists of two perspectives: (1) testing changes in behaviors of skills used across time (pre-intervention vs. post-intervention) and between intervention groups (P vs. PF); and (2) analyzing the relationship between intervention-exposure to good alternative patterns in AI feedback and post-intervention skill use.

\paragraph{Testing Changes Across Time and Between Groups}
Using our selected classifiers, we examined skill usage changes from pre- to post-intervention. For each transcript, we computed the proportion of utterances showing strong skill use ($b = U_{\text{strengths}} / U_{\text{total}}$) or needing improvement ($b = U_{\text{improvement}} / U_{\text{total}}$).

To test for pre-post changes ($b_0, b_1$), we used paired $t$-tests and Cohen's $d$ effect sizes. To compare P and P+F groups ($b_1^P-b_0^P$ vs. $b_1^{PF}-b_0^{PF}$), we used unpaired $t$-tests. We conducted 12 planned $t$-tests: three skills (Empathy, Reflections, Questions) for strong uses and one skill (Suggestions) for areas needing improvement, analyzing both within-group changes and between-group differences.


\paragraph{Exposure to Good Alternatives in AI Feedback}
To assess whether AI feedback exposure affected post-intervention performance, we defined Good Alternatives during Practice (GAP) as the proportion of trainee utterances for which the AI suggested an alternative response that exemplified a strong use of a skill:
\[
T_{GAP} = \frac{A_{strengths}}{U_{total}},
\]
where \(A_{strengths}\) is the number of AI-generated alternative responses judged as strong exemplars and \(U_{total}\) is the total trainee utterances. We then fit a lagged linear regression predicting post-chat behavior \(b_1\) while controlling for pre-chat behavior \(b_0\) and including GAP as a predictor:
\[
b_1 = \beta_0 + \beta_1 b_0 + \beta_2 T_{GAP},
\]
thereby isolating the effect of feedback exposure.

\subsubsection{\textbf{RQ2. Effects on Self-Efficacy and its (mis)calibration with Behavioral Performance}}

First, we test for changes in raw self-efficacy scores after practice or practice-and-feedback. Second, we examine the calibration of self-efficacy ratings with actual performance. Third, we evaluate whether P or P+F interventions improve this calibration.

\paragraph{Changes in Raw Self-Efficacy}
Beyond calibration, we also investigated whether the interventions affected participants' absolute levels of self-efficacy across the three measured dimensions (exploration skills, action skills, and session management skills). We conducted repeated-measures analyses to identify: (1) significant pre-post changes in raw self-efficacy scores following practice alone (P intervention); (2) significant pre-post changes in raw self-efficacy scores following practice with structured feedback (P+F intervention); and (3) differential patterns of change between the P and P+F groups, indicating potential intervention-specific effects on self-efficacy development.

\paragraph{Investigating (mis)calibration of Self-Efficacy}
Our primary analysis investigated potential mis-calibration between participants' self-assessments and their actual counseling performance, specifically examining whether data exhibited patterns consistent with the Dunning-Kruger effect. This phenomenon~\cite{kruger1999unskilled} suggests that individuals with lower skill levels tend to overestimate their abilities, while highly skilled individuals may slightly underestimate their competence.
We focused this analysis on Exploration Skills and Action Skills, as these dimensions had straightforward mappings between CASES items and our NLP behavioral classifiers (Table~\ref{tab:mapping_perceived_actual_ability}).
To test for the presence of the Dunning-Kruger effect, we follow the classic analysis method that splits the data into quartiles based on performance and conducts a two-way analysis of variances for self-assessments and actual performance across the quartiles; and finally verifies via post-hoc tests that the bottom performers have the biggest overestimation of their abilities~\cite{kruger1999unskilled}. To standardize the comparison between self-efficacy and performance, we transform each measure into a percentile rank (0 - 99) computed across all data collected for the pre-intervention and post-intervention chats ($b_0, b_1$; $s_0, s_1$).

\paragraph{Changes in Calibrated Self-Efficacy}
To evaluate whether our interventions improved self-efficacy calibration, we computed discrepancy scores by subtracting standardized performance scores from standardized self-efficacy scores for each participant at both assessment timepoints. These discrepancy metrics provided a direct measure of calibration, with positive values indicating overconfidence and negative values indicating underconfidence.
We then examined changes in these discrepancy scores from pre- to post-intervention for both intervention groups, to determine whether either intervention improved the alignment between participants' self-perceptions and their actual counseling abilities.

\subsubsection{\textbf{RQ3. Effects on Therapeutic Intentions}}
Three authors conducted a thematic analysis~\cite{braun2006thematicanalysis} of participants' post-intervention reflections on "what they would do differently." To make coding of 94 transcripts feasible, authors used timestamped session notes to locate and extract relevant transcript excerpts. We started with codes derived from the Helping Skills taxonomy (Table~\ref{tab:skill_descriptions}) and then inductively generated the following codes: Empathy, Validation, Action Plan, Active Listening, Questions / Open-Ended, Suggestions, Trust/Connection, Confidence / Personal Growth, Reframing / Affirmations, Reflection, Self-Disclosure, Professionalism, Personalization, and Nothing to improve. Three co-authors independently coded the data; the group compared coding and disagreements were resolved via iterative discussion. Condition-specific frequency tables are provided in Appendix~\ref{tab:skill_freq_pf} and~\ref{tab:skill_freq_p}. These codes were synthesized into higher-level themes informed by literature on therapeutic intentions and microskills~\cite{hill2000client, raskin2005person}.

\subsubsection{\textbf{RQ4. Quantitative Perceptions of Receiving LLM Feedback and Practicing with Simulated Patients}}
For the Likert survey questions, we report descriptive statistics of all Likert measures that capture participants' perceptions of \system. Consistent with recent papers analyzing the convergent validity of the NASA-TLX instrument in HCI~\cite{babaei2025should}, we consider it as a multivariate construct in our analysis.

\subsubsection{\textbf{RQ5 and RQ6. Qualitative Experience Receiving LLM Feedback and Practicing with Simulated Patients}}

To understand how participants experienced both \system's feedback and simulated patients, we conducted a thematic analysis~\cite{braun2006thematicanalysis} of intervention session recordings with P+F participants. To analyze the variation in participants' experience with \system feedback, we used purposeful sampling~\cite{palinkas2015purposeful} based on the intensity of participants' survey responses to the feedback helpfulness item (``To what extent do you find the AI feedback to be constructive and helpful?''). From the P+F condition, we selected 16 participants ($\sim$33\% of the sample): eight who rated feedback helpfulness as moderate or lower ($\leq$3 on a 5-point scale) and eight who rated it as high ($\geq$4), ensuring representation of both positive and critical perspectives. For each sampled participant, the first author analyzed P+F participants' Zoom recording, with particular attention to commentary during the intervention chat when \system's feedback system was used; the post-intervention chat in which participants had a chance to continue or shift their approach based on what they learned in the intervention chat; and exit-interview responses. The transcript was coded using a deductive set of high-level organizing categories---negotiating with and integrating the AI feedback and perceptions of simulated conversations with AI patients---and inductively coded within each category. When participants referenced specific AI feedback during think-alouds, the corresponding dialogue history and feedback content was exported from the \system web platform and linked to their commentary to contextualize their thoughts.

%% file: sections/5_results.tex
\section{Results}\label{sec2}

\subsection{RQ1. Effects on Behavioral Performance} 


\begin{figure*}[t]
    \centering
    {\includegraphics[width=\textwidth]{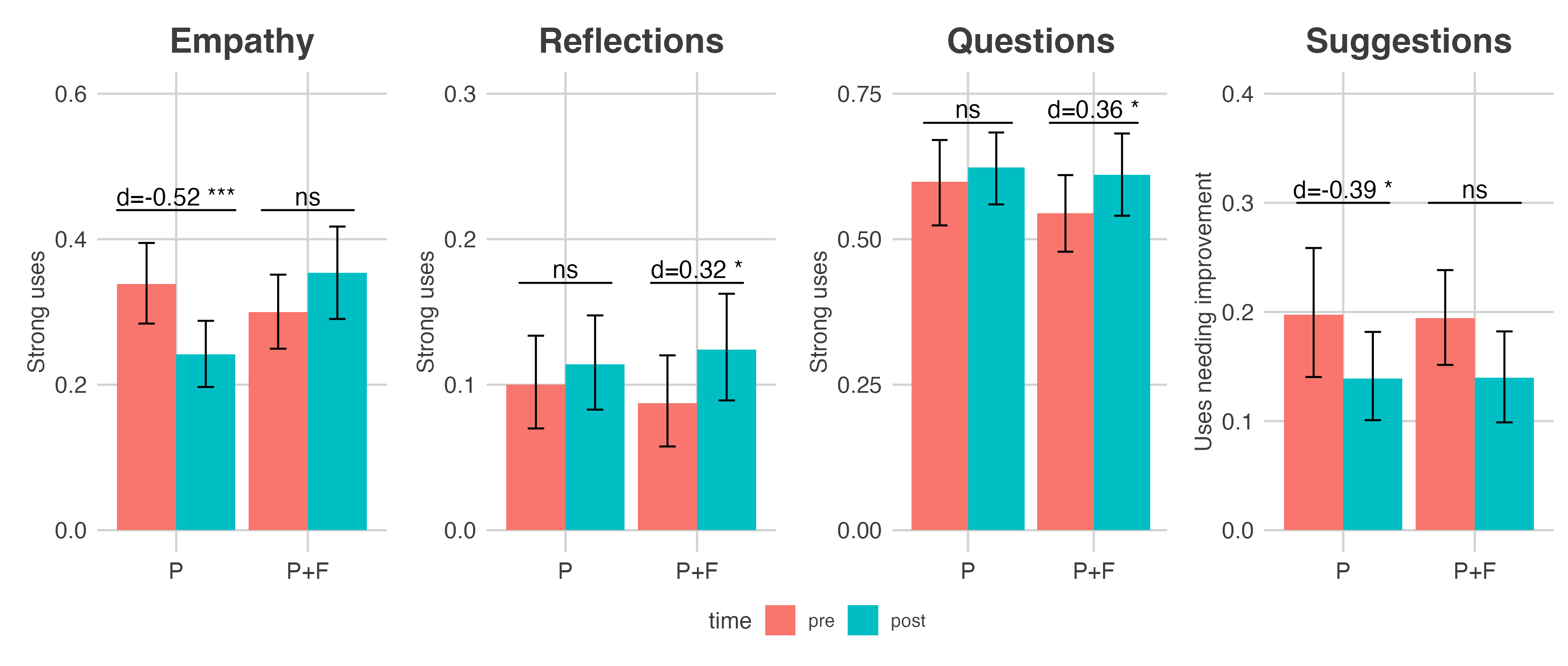}}
    \resizebox{\textwidth}{!}{
    \begin{tabular}{l c c c c c c c c c}
    \toprule
     & \multicolumn{3}{c}{Change after P} & \multicolumn{3}{c}{Change after PF} & \multicolumn{3}{c}{Differences in} \\
     & \multicolumn{3}{c}{} & \multicolumn{3}{c}{} & \multicolumn{3}{c}{change after P vs. P+F} \\
    \cline{2-10}
    \textbf{Use of Skill} & \% & \textit{p}-value & \textit{d} & \% & \textit{p}-value & \textit{d} & \% & \textit{p}-value & \textit{d} \\
    \midrule
    Empathy ($\uparrow$) & \textbf{-9.6} & \textbf{0.001} & \textbf{-0.52} & 5.4 & 0.117 & 0.23 & \textbf{15} & \textbf{0.001} & \textbf{0.72} \\
    Reflections ($\uparrow$) & 1.4 & 0.391 & 0.18 & 3.7 & 0.034 & 0.32 & 2.3 & 0.323 & 0.2 \\
    Questions ($\uparrow$) & 2.4 & 0.421 & 0.12 & 6.6 & 0.018 & 0.36 & 4.1 & 0.296 & 0.22 \\
    Suggestions ($\downarrow$) & \textbf{-5.9} & \textbf{0.010} & \textbf{-0.39} & -5.5 & 0.057 & -0.28 & 0.4 & 0.910 & 0.02 \\
    \bottomrule
    \end{tabular}
    }
    \caption{\small{Changes in counseling behaviors following AI patient simulations alone (P) versus AI patient simulations with AI feedback (P+F). The plot displays bootstrapped means for pre-intervention and post-intervention interactions. The table presents statistical comparisons with corresponding effect sizes, with \textbf{bolded} values indicating significance after Benjamini-Hochberg correction~\cite{benjamini1995controlling}. 
    Notably, the P group experiences a significant drop in strong uses of Empathy (-9.6\% change, $d=-0.52$), whereas the P+F group's use of Empathy is maintained and trends towards improvement; the large between-group difference (15\% difference, $d=0.72$) indicates the causal impact of feedback.  Conversely, while the P group had fewer inappropriate uses of Suggestions (-5.9\% change, d = -0.39), the between-group difference is close to zero (0.4\% difference, $d=0.02$), indicating that another mechanism besides feedback is driving this change. The P+F group also experiences noticeable improvements in Reflections (+3.7\% change, $d=0.32$) and Questions (6.59\% change, $d=0.36$)
    }}
    \label{fig:mainfig_goodareas}
    \Description{The figure shows four side-by-side bar plots illustrating changes in counseling behaviors pre-(red bars) and post-(blue bars) intervention, comparing the Practice-only (P) group and the Practice+Feedback (P+F) group. Empathy panel: In the P group, strong uses of empathy significantly decreased from pre- to post-intervention (Cohen’s d = –0.52, p < .001). In the P+F group, empathy remained stable with no significant change. Between-group comparison shows that P+F participants expressed more empathy than P participants after training. Reflections panel: In the P group, reflections did not significantly change. In the P+F group, strong uses of reflections significantly increased (d = 0.32, p < .05). Questions panel: In the P group, there was no significant change. In the P+F group, strong uses of questions significantly increased (d = 0.36, p < .05). Suggestions panel: In the P group, suggestions needing improvement significantly decreased (d = –0.39, p < .05). In the P+F group, there was no significant change. Overall, the results show that AI feedback (P+F) helped improve reflections and questions while preventing empathy decline, whereas practice alone (P) reduced empathy but lowered poor suggestions.}
\end{figure*}

\textbf{We find that practice alone is not enough; feedback during practice is necessary to promote desirable counseling behaviors in empathetic and active listening.} AI feedback during practice (P+F) led to improvements in Reflections (+3.6\% change, $p=0.034$) and Questions (+6.59\% change, $p=0.018$), and trended toward improvement in 
Suggestions (-5.45\% change, $p=0.057$) and Empathy (+5.37\% change, $p=0.117$). In contrast, practice alone (P) showed a different pattern: while participants reduced inappropriate Suggestions (-5.85\% change, $p=0.011$), they significantly {\em worsened} in Empathy (-9.6\% change, $p<0.001$), with no improvements in Reflections or Questions.

Between-group comparisons of skill change allowed for estimating the effect of \system{}'s feedback. Empathy showed a substantial and significant difference  P+F (15\% relative difference, Cohen's $d=0.72, p<0.001$), indicating a large feedback effect. Conversely, Suggestions showed near-zero between-group differences ($d=0.02, p=0.910$) despite pre-post improvement, suggesting that another mechanism besides feedback is driving the reduction in inappropriate suggestions.

To better understand AI feedback's role, we further analyzed how behavioral performance is impacted by exposure to specific feedback during the practice-intervention. 
\textbf{For empathy skills, exposure to alternatives with strong uses of empathy during training significantly predicted post-intervention empathy scores ($\beta_2 = 0.204, p = .018$).} 
However, exposure to good alternatives did not significantly predict improvement in other counseling skills (Reflections: $\beta_2 = 0.049, p = .440$; Questions: $\beta_2 = 0.046, p = .523$). This suggests that the effectiveness of AI feedback alternatives varies by skill type, with empathy skills appearing more responsive than reflections or question skills. 

\begin{figure*}[t]
    \centering
    {\includegraphics[width=\linewidth]{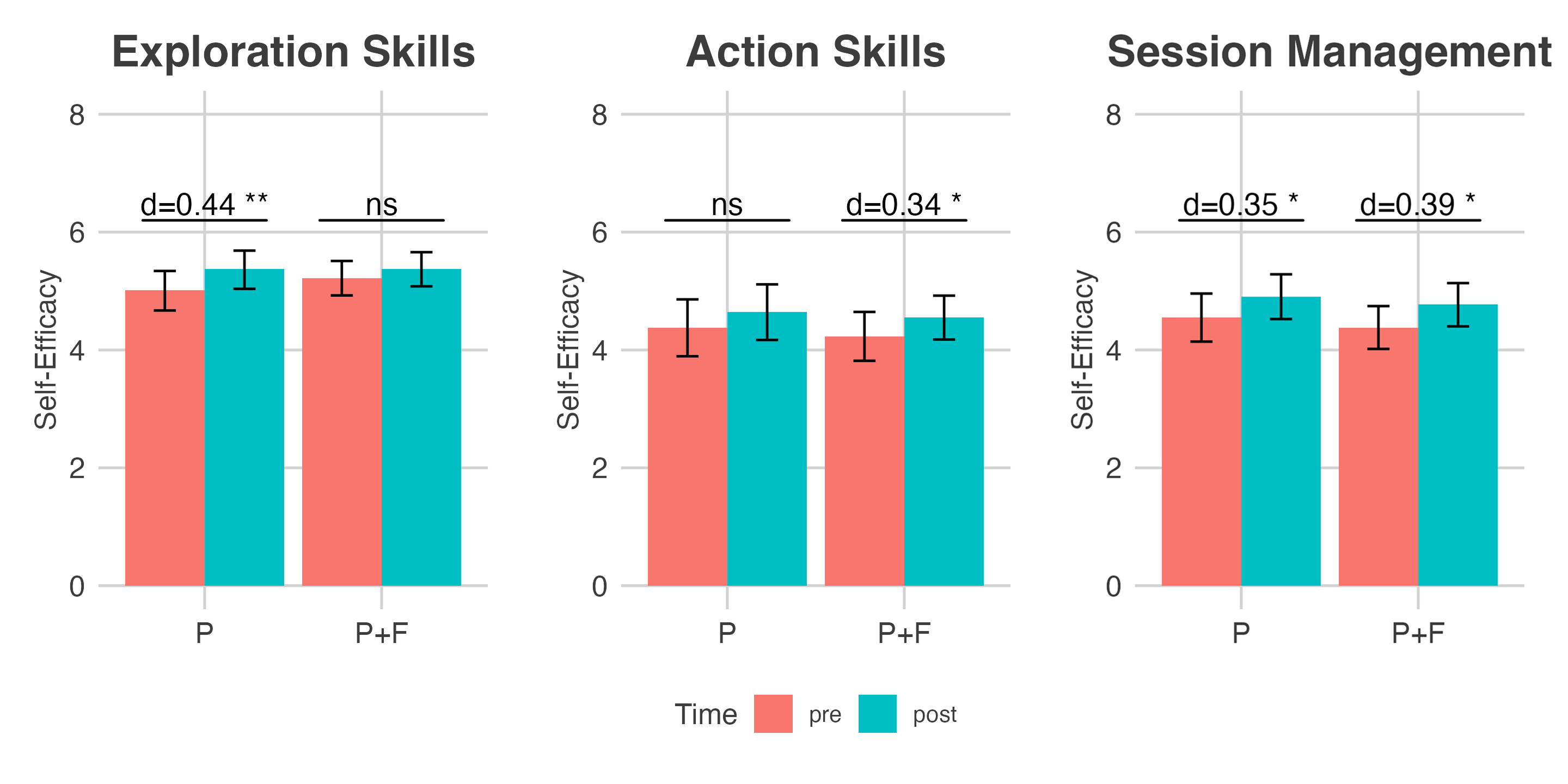}}
    \resizebox{\linewidth}{!}{
    \begin{tabular}{l c c c c c c c c c}
    \toprule
     & \multicolumn{3}{c}{Change after P} & \multicolumn{3}{c}{Change after PF} & \multicolumn{3}{c}{Differences in} \\
     & \multicolumn{3}{c}{} & \multicolumn{3}{c}{} & \multicolumn{3}{c}{change after P vs. PF} \\
    \cline{2-10}
    \textbf{Self-Efficacy} & Pts. & \textit{p}-value & \textit{d} & Pts. & \textit{p}-value & \textit{d} & Pts. & \textit{p}-value & \textit{d} \\
    \midrule
    Exploration Skills & \textbf{0.36} & \textbf{0.004} & \textbf{0.44} & 0.13 & 0.338 & 0.14 & -0.21 & 0.238 & -0.25 \\
    Action Skills & 0.27 & 0.166 & 0.21 & 0.33 & 0.026 & 0.34 & 0.04 & 0.884 & 0.03 \\
    Session Management & 0.35 & 0.021 & 0.35 & \textbf{0.36} & \textbf{0.011} & \textbf{0.39} & 0.01 & 0.955 & 0.01 \\
    \bottomrule
    \end{tabular}
    }
    \caption{\small{Changes in raw-scores of self-efficacy following AI patient simulations alone (P) versus AI patient simulations with AI feedback (PF). The plot displays bootstrapped means for pre-intervention and post-intervention. The table presents statistical comparisons with corresponding effect sizes, with \textbf{bolded} values indicating significance after Benjamini-Hochberg correction~\cite{benjamini1995controlling} for the 21 planned comparisons (12 for behavioral changes and 9 for self-efficacy changes).}}
    \label{fig:selfeff_means}
    \Description{The figure presents three side-by-side bar plots showing changes in self-efficacy scores pre- (red bars) and post- (blue bars) intervention, comparing the Practice-only (P) and Practice+Feedback (P+F) groups. Exploration Skills panel: The P group showed a significant increase in exploration self-efficacy (Cohen’s d = 0.44, p < .01), while the P+F group showed no significant change. Action Skills panel: The P group showed no significant change. The P+F group significantly increased in action skills self-efficacy (d = 0.34, p < .05). Session Management panel: Both groups showed significant increases in self-efficacy for session management (P: d = 0.35, p < .05; P+F: d = 0.39, p < .05). Overall, participants reported modest improvements in self-efficacy across skill domains, but changes did not consistently align with actual performance improvements shown in Figure 3.}
\end{figure*}

\subsection{RQ2. Self-Efficacy and Its Miscalibration with Behaviors}
\textbf{In our analysis of raw self-efficacy scores, we find modest overall increases in self-efficacy after P and P+F interventions, with different patterns of improvement across skills} (Fig.~\ref{fig:selfeff_means}).  For the P group, confidence in exploration skills showed a significant increase ($0.36$ points on an 8-point scale, $d=0.44, p=0.004$). Confidence in session management skills showed a substantial increase for the P+F group ($0.36$ points, $d=0.39, p=0.011$). While session management skills for the P group also trended towards improvement, it was not significant after correcting for multiple hypothesis testing ($0.35$ points, $d=0.35,p=0.021$). Similarly, while confidence in action skills for the P+F group also increased, this result was not significant after correction of multiple hypothesis tests ($0.33$ points, $d=0.34, p=0.026$).  Finally, we found no significant differences between participants who received AI feedback (P+F) versus those who did not (P) (across the three self-efficacy subscales, $d=-0.25, 0.03, 0.01$, $p=0.238, 0.884, 0.955$).

\begin{figure*}[t]
    \centering
    \includegraphics[width=\linewidth]{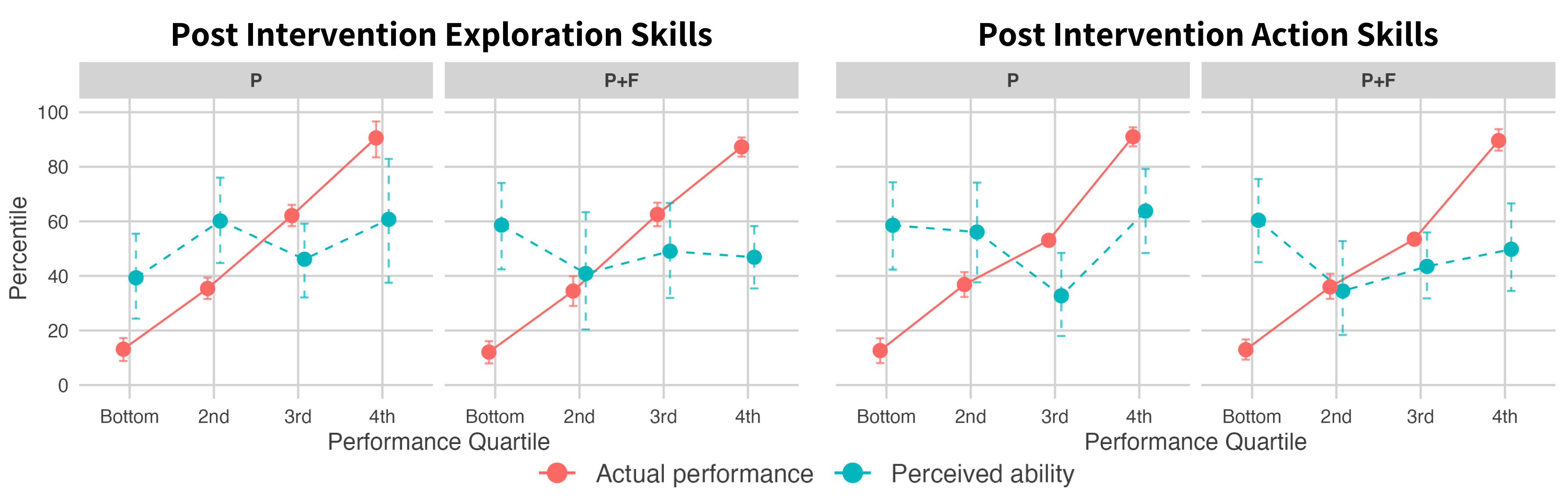}
    \caption{\small{Counselor Self Efficacy (perceived ability to use skills) for participants grouped by behaviors of skills used (actual performance). Notes: Gaps depict miscalibration between actual and self-assessed percentile of performance for quartile groups with bootstrapped 95\% CIs. We only visualize data collected in the post-intervention.
    }}
    \label{fig:dunning_kruger_comparison}
    \Description{The figure shows line plots comparing actual performance (red solid line with circles) and perceived ability (blue dashed line with circles) across quartiles of counselor performance after training. Each panel shows a different skill domain. Exploration Skills: In both the Practice-only (P) and Practice+Feedback (P+F) groups, actual performance increases steadily from the bottom to the top quartile. In contrast, perceived ability stays relatively flat, with the lowest quartile overestimating their skill and the highest quartile slightly underestimating it. Action Skills: A similar miscalibration pattern is seen. Actual performance rises sharply across quartiles, while perceived ability fluctuates and does not track performance. The lowest performers show the largest overestimation. Error bars represent bootstrapped 95\% confidence intervals. The gaps between red and blue lines depict miscalibration, consistent with Dunning-Kruger effects: bottom-quartile participants overestimate their ability most, while top-quartile participants modestly underestimate themselves.}
\end{figure*}

\textbf{Our analysis comparing self-efficacy ratings with actual performance across skill quartiles finds support for the Dunning Kruger effects more substantially for action skills and to a lesser degree for exploration skills.} The interaction between measure and quartile was significant in four out of six ANOVAs for action skills, while only one out of six was significant for exploration skills (Table~\ref{tab:dke_ESAS_anova}). Pairwise comparisons also showed a pattern indicative of a Dunning-Kruger effect (see Table~\ref{tab:dke_pairwise_timepoint_quartile}, Table~\ref{tab:dke_pairwise_timepoint_group_quartile}, and Fig.~\ref{fig:dunning_kruger_comparison}): People in the lowest quartile overestimated themselves the most. 
Those in the highest quartile—and to a lesser degree also those in the second-to-highest quartile—tended to underestimate themselves.



\textbf{Participants' Ability to Self-Assess Their Skill Level Remained Mixed After LLM Practice}.
For the practice only (P) group, the mean discrepancy in exploration skills changes from 11.6 percentile underconfidence to 5.7 percentile overconfidence, a significant shift ($p<0.001, d=0.58$). 
Besides this, we found no other significant changes in calibration. No significant differences was found among the P group for discrepancy in action skills ($p=0.227, d=0.18$). The P+F group showed no significant calibration changes for exploration skills ($p=0.479, d=-0.10$) or for action skills ($p=0.393, d=0.13$). Finally, between-group differences were not significant for discrepancy in exploration skills ($p=0.191, d=0.27$) or action skills ($p=0.743, d=0.07$).

\subsection{RQ3. Qualitative Self-Reflections on Therapeutic Intentions}

Two key themes emerged for how novice counselor's therapeutic intentions were impacted by training with \system:

\textbf{(1) The P+F group expressed greater intentions and successes in improving their use of empathy and listening skills.} 
P+F participants reported effectively using empathy (27\%), validation (27\%), and open-ended questions (52\%). They emphasized the value of listening skills, such as reflective responses to signal understanding: \textit{``I should rephrase what they say so they know I’m understanding them''} (P51). They also recognized that counseling should support client exploration of thoughts and emotions, rather than provide direct solutions. One participant reflected on this shift: \textit{``I asked them to expand on their feelings, rather than guiding them to my idea''} (P39). AI feedback encouraged this shift toward providing emotional support and fostering client autonomy, helping participants adopt a more empathetic and client-centered approach.
\textbf{(2) The P participants remained solution-oriented but changed their approach to first gather information.} While P+F participants intended to give fewer suggestions, many P participants continued to view suggestions as a central skill.
In the post-intervention, 48\% of P participants reported using suggestions successfully, compared to only 14\% of P+F participants. Many P participants justified their continued use of suggestions by citing a desire to provide tangible, actionable help, for example, \textit{``Maybe because I'm untrained and solution oriented. I do not want to leave them with nothing, and nowhere to go''} (P40). Some reflected on modifying how they delivered suggestions, emphasizing strategies like gathering more information to tailor advice or providing more concrete guidance. However, in the absence of feedback, most remained fixed in their approach, with some even reporting efforts to rephrase the same solution repeatedly to persuade the client.

\subsection{RQ4. Quantitative Experience of Receiving LLM Feedback and Practicing with Simulated Patients}

\subsubsection{Quantitative Perceptions of \system Feedback}
\textbf{Participants in our study had consistently positive perceptions of \system's generated feedback across multiple dimensions.} The majority of participants (76\%) found the AI feedback constructive and helpful ($\mu = 4.1$ out of 5, $\sigma = 0.8$), while an even higher proportion (84\%) reported being comfortable receiving feedback from the AI system ($\mu = 4.4$ out of 5, $\sigma = 0.9$). Additionally, 72\% agreed that the AI feedback system is ready for use by counselors-in-training ($\mu = 3.8$ out of 5, $\sigma = 1.0$). These consistently high ratings across helpfulness, comfort, and readiness measures suggest strong overall acceptance of AI-generated feedback among participants. 

\subsubsection{Quantitative Perceptions of Training with AI Patients}
Descriptive statistics of participants' perceptions are shown in Table~\ref{tab:ai-patient-perceptions}.
\textbf{Most participants felt that AI patients in \system were realistic.} Across all three AI patient scenarios, the vast majority of participants (88--92 out of 94) rated the AI patients as authentic in their roles, with scores of 5, 6, or 7 on the 7-point Likert scale. Authenticity ratings were consistently high across scenarios ($\mu = 6.1$--$6.3$, $\sigma = 0.9$--$1.0$). 
\textbf{Participants consistently found the AI patient most challenging during the intervention phase across multiple measures.} Mental demand peaked during intervention, with 82\% of participants rating it as moderate to high ($\mu = 5.5$, $\sigma = 1.2$) compared to 68\% pre-intervention ($\mu = 4.7$, $\sigma = 1.4$) and 56\% post-intervention ($\mu = 4.5$, $\sigma = 1.4$). Similarly, perceived effort required was highest during intervention, with 81\% rating it as moderate to high ($\mu = 5.4$, $\sigma = 1.3$) versus 64\% pre-intervention ($\mu = 4.8$, $\sigma = 1.4$) and 65\% post-intervention ($\mu = 4.8$, $\sigma = 1.3$). Frustration levels, while lower overall, also peaked during intervention with 40\% reporting moderate to high frustration ($\mu = 3.9$, $\sigma = 1.4$) compared to 35\% pre-intervention ($\mu = 3.6$, $\sigma = 1.5$) and 30\% post-intervention ($\mu = 3.5$, $\sigma = 1.6$). This consistent pattern across these NASA-TLX measures suggests that either the extended intervention practice imposed the greatest demands on participants, or that the intervention simulated patient was the most challenging scenario for our subject pool of participants.
\textbf{Participants felt progressively less hurried or rushed across intervention phases.} 45\% (42/94) rated feeling moderately to highly hurried or rushed ($\geq$5 on Temporal Demand) during pre-intervention ($\mu = 3.9$, $\sigma = 1.8$), compared to 31\% (29/94) during the intervention phase ($\mu = 3.6$, $\sigma = 1.7$) and 30\% (28/94) in the post-intervention phase ($\mu = 3.3$, $\sigma = 1.8$). This decreasing trend suggests that participants became more comfortable with the pacing of AI patient interactions over time. 

\begin{table*}[t]
\centering
\begin{tabular}{lccccccc}
\hline
& \multicolumn{2}{c}{Pre-Intervention} & \multicolumn{2}{c}{Intervention} & \multicolumn{2}{c}{Post-Intervention} \\
\cmidrule(lr){2-3} \cmidrule(lr){4-5} \cmidrule(lr){6-7}
Measure & $\mu$ & $\sigma$ & $\mu$ & $\sigma$ & $\mu$ & $\sigma$ \\
\hline
Authentic in role & 6.1 & 1.0 & 6.2 & 1.0 & 6.3 & 0.9 \\
Mental Demand & 4.7 & 1.4 & 5.5 & 1.2 & 4.5 & 1.4 \\
Temporal Demand & 3.9 & 1.8 & 3.6 & 1.7 & 3.3 & 1.8 \\
Effort & 4.8 & 1.4 & 5.4 & 1.3 & 4.8 & 1.3 \\
Frustration & 3.6 & 1.5 & 3.9 & 1.4 & 3.5 & 1.6 \\
Confidence to help & 4.8 & 1.3 & 4.8 & 1.4 & 5.3 & 1.4 \\
\hline
\end{tabular}
\caption{Perceptions of training with \system's AI patients across the three study phases. NASA-TLX dimensions including Mental Demand, Temporal Demand, Effort, and Frustration highlight the experience trying to interact and provide counseling support to the patient. AI Patient 1 (Pre-Intervention) was the 35-year-old American Male who was feeling alone after a holiday; AI Patient 2 (Intervention) was the 35-year-old Male Veteran who had substance use and legal issues retaining custody of his kids; AI Patient 3 (Post-Intervention) was the young adult with family issues who had low mood and self-esteem. 7 Point Likert Scale. 
}
\label{tab:ai-patient-perceptions}
\end{table*}

\subsection{RQ5. Qualitative Perceptions of Receiving LLM Feedback}\label{sec:qual-feedback}
In this subsection, we describe how participants productively used the AI feedback, ways they negotiated whether to accept or reject the AI feedback's suggestions, and how receiving AI feedback affected their self-confidence and sense of professional identity.

\subsubsection{Productive Uses of AI Feedback}
Participants described several ways they found the AI feedback helpful for developing their counseling skills: generating alternative phrasings and conversational directions and checking their intentions in real-time.

\textbf{Generating Alternative Phrasings and Conversation Directions.}
When participants aimed for empathetic listening, the AI's alternatives helped them tighten phrasing and redirect stalled exchanges. For example, one noted \textit{"Alternative responses helped with validation... being supportive and addressing concerns before moving on"} (P107). Another swapped \texttt{"Have [your parents] seen your efforts...?" }for \texttt{"Can you tell me more about how you have been trying to show [your parents]...?"} and said \textit{"Oh, that is a good idea... it felt like it would make the patient think more"} (P75). Others used alternatives to break conversational loops: \textit{"I was getting stuck... the feedback directed me to ask 'Can you tell me more...?' which I would have missed"} (P91). Overall, alternatives helped participants sound more supportive, elicit deeper exploration, or change the conversation's direction.


\textbf{Real-Time Intention Checking and Self-Reflection.} Many participants reported using the AI feedback as an on-demand check to confirm that their intended therapeutic stance was being communicated. For example, one participant described it as helping them verify their tone: \textit{``...review that feedback in real time and see, `Yes, this still does sound like an empathetic response'...''} (P55). Several others echoed that the strengths portion was particularly reassuring: \textit{``...it reinforced that I was on the right track...''} (P65). More broadly, participants emphasized that pausing to read feedback prompted reflection about their approach. As one put it, \textit{``It helps you to reflect on what you are saying to people; even if you don't agree''} (P24B). A smaller subset who checked feedback infrequently noted regret and wished they had engaged more: \textit{``I could have course corrected earlier...''} (P17).

\subsubsection{Negotiating Whether to Accept or Reject AI Feedback}
While participants found value in the AI feedback, they did not accept it uncritically. Our analysis revealed several factors that influenced whether participants integrated or dismissed the feedback: logical consistency with the dialogue context, alignment with their clinical goals and intuitions, perceived appropriateness for the specific scenario, and ecological validity for real-world practice.

\textbf{Rejecting Feedback That Was Logically Inconsistent with the Dialogue.} Several participants reported dismissing AI suggestions that failed to track prior disclosures or that simply restated content already covered. For example, one participant noted: \textit{``This one, it kind of said that I was assuming things about the seeker's feelings? But they kind of outright told me that. Their parents keep bringing up what happened 6 months ago. So it wasn't really an assumption''} (P77). Others rejected feedback that echoed their own prior wording: \textit{``It's exactly the same thing I've said... just different words''} (P89). A number of participants also pointed out internal contradictions across sequential feedback items: \textit{``...it doesn't want me to ask an open-ended question, and then the next feedback... is what it just told me not to do''} (P97).

\textbf{Protecting Clinical Judgment and Strategic Goals.} Several participants described rejecting feedback when it conflicted with deliberate clinical choices or situational judgment. They framed some phrasing choices as strategic attempts to elicit disclosure rather than leading the patient, and defended clarifying questions as necessary when context was incomplete. One participant explained that an indirect phrasing was intentional: ``The reason I said `a trip, like a vacation?' is I didn't want to be too forward ... I wanted him to bring up the drugs'' (P45). The same participant emphasized the need to gather context before offering guidance: ``I would disagree ... if I don't know the whole picture, I can't even give vague enough stuff ... there's an amount of context that it's alright to ask for'' (P45).
Others used positive responses from patients as justification to keep their original approach. For example, one participant reported that a suggestion to ``speak with a religious leader'' had elicited an open response from the patient, so they retained that line of inquiry (P85). This suggests that conflicting signals between the feedback and the patients leaves it to the counselor to decide.

\textbf{Domain expertise overriding AI recommendations.} Experienced participants sometimes rejected AI advice as inappropriate for high-risk cases. For example, a counselor with substance-use experience summarized:
\begin{quote}
\textit{"...They're good at first... then it goes into more empathy and lacks understanding of addiction cases... concern should shift to the family... I'd be calling Child Protection Services."} (P33)
\end{quote}
They called the AI's empathy-focused responses "a little scary" and said they "would never want a machine to deal with somebody like that" (P33). Empathy and validation can generally be good to use, but as this participant highlights, the such empathy is not warranted if its neglected safety considerations in high-risk scenarios.

\textbf{Questioning the Repetitive Emphasis on Empathy and Validation.} Several participants felt the feedback over-emphasized empathetic reflections, to the point of seeming unrealistic: \textit{"it keeps saying... 'you need to regurgitate their feelings'... I don't know if that's realistic"} (P79). Others observed that corrections repeatedly \textit{"focused on feelings and emotions"} rather than practical help (P91). Participants generally resolved this by noting context matters: validation may be appropriate for some patients (e.g., resistant cases), while more practical responses suit others (\textit{"this one is willing to work on the practical..."} (P45); \textit{"maybe... good for resistant patients"} (P97)).

\textbf{Ecological Validity Concerns.} Participants also evaluated feedback against the practical constraints of real-world counseling. Several noted that the AI's suggested responses were unrealistically lengthy for real-time text-based interaction: \textit{``What they're suggesting is a lot longer and more in-depth response than what I put... there is a speed of how quick I can respond as opposed to AI''} (P95). The same participant noted that the AI did not account for session length: \textit{``If you've got a 2 hour session planned, then you can keep summarizing and allow somebody a lot more space and time to dig a bit deeper. What if you've got a 10 minute interaction? Then I think it needs to be a bit more concise''} (P95).

\subsubsection{Impacts on Self-Confidence and Professional Identity}
Beyond the content of the feedback itself, participants described how the experience of receiving AI feedback affected their emotional state, self-confidence, and sense of professional identity. While some found the feedback encouraging, others experienced it as demoralizing or as a threat to their authentic voice as a counselor.

\textbf{Discouragement from Pervasive Criticism.} Participants receiving feedback on most responses felt overwhelmed. One participant who received suggestions on six of seven responses reflected: \textit{``Not great... everything was challenged. So I almost feel like I'm not doing very well...''} (P91). They elaborated: \textit{``Each response offers a better way. It feels like nothing's ever going to be right for this AI''} (P91). The frequency of critical feedback shaped emotional responses more than content alone. Repetitive corrections on the same issue frustrated participants: \textit{``The AI feedback kept saying the same thing, and that made me feel upset''} (P89).

\textbf{Threat to Authenticity and Professional Identity.} Some participants worried that using AI suggestions might erode their authentic counseling voice. One participant noted a shift from openness to resistance:
\begin{quote}
\textit{``[My original response] was a complex reflection which I thought was fine... After reading [the AI's alternative], I questioned myself since the AI's were more well-rounded. But still like that is me, right? I guess that's how I [respond]. So I don't want the AI to take me away from the sessions... I'll become like a generic text counselor, or off the shelf counselor.''} (P91)
\end{quote}
This tension between improvement and authenticity highlights concerns about whether modeling the AI's alternatives signifies skill development or a loss of individual therapeutic style.

\subsection{RQ6. Qualitative Perceptions of Practicing with LLM-Simulated Patients}\label{sec:qual-patients}

In this subsection, we describe how participants experienced practicing with AI-simulated patients, focusing on their reactions to resistant patient behaviors and how personal relatability with simulated patients' identities influenced the practice experience.

\textbf{Participants had divergent reactions to AI patients' resistant behaviors.} The AI patients were designed to resist advice and suggested actions, which participants consistently noticed. Some welcomed this resistance as valuable preparation for real clinical work: \textit{"There's almost like a stubbornness to them... And I think that's good, especially for people that don't have any experience in counseling... I think [CARE] is a better way to ease into working with a resistant patient"}(P97).
However, others experienced discouragement when their efforts to support the patient were repeatedly deflected: \textit{"They were very cold, it was hard to communicate with this person... I'm a tiny bit discouraged in myself, the patient was not taking what I was suggesting very well"} (P44B).

Some participants also questioned the authenticity of the resistance patterns. One noted that the AI seemed \textit{"in a loop"} of refusal that felt unlike real human behavior: \textit{"It feels like the [simulated] person has a trained response to basically refuse any care suggestion, but in the nicest way possible. Most [real] people just lie. So after the 3rd or 4th question... most people would say 'Oh, that's a really good idea. I'll do that' just to get you to shut up... So that's where it kinda falls apart"} (P33). This suggests that while consistent resistance provides useful challenge, calibrating the degree and style of resistance to feel more naturalistic remains an area for refinement. These divergent reactions underscore that participants' individual readiness to handle difficult cases directly affects whether they perceive simulations as valuable learning experiences.

\textbf{Personal relatability with simulated patients influenced perceived difficulty and engagement.} Several participants reported that variation in age, gender, and presenting concerns required them to adapt their responses; encountering demographically or experientially dissimilar patients often felt harder: \textit{"...white, older, middle-aged males, who had kids... I couldn't relate"} (P75). Some presenting scenarios were more difficult when a novice counselor had a hard time relating to their experience: \textit{"I'm really close to my family, whereas they are estranged from theirs, so I just felt kind of stuck as to what to say or suggest"} (P41). By contrast, perceived similarity tended to boost confidence and connection: \textit{"If a woman talks to a woman... they can relate more... I was already achieving the goal"} (P89), and allowed some participants to draw on lived experience: \textit{"I could definitely empathize... I've been there"} (P75). Together these accounts imply that exposing novices to dissimilar patients may foster broader preparation for scenarios, but such practice can be discouraging without added scaffolding (e.g., brief prompts, reflection questions, or supervisor guidance). 

\textbf{The rapid pace of AI responses created artificial time pressure that affected practice quality.} Several participants described feeling rushed because AI patients replied almost instantly, which increased temporal demand and sometimes disrupted clinical attentiveness. For example, one participant summarized this experience as \textit{``...a speed of how quick I can respond as opposed to AI...''} (P95), and another said \textit{``They respond so quickly... you feel a kind of pressure to respond back''} (P22). A number of participants linked this perceived rush to concrete interaction problems: \textit{``...I was kind of asking some of the same questions because I felt a little rushed... he mentioned that the therapy was court ordered earlier, but I asked him the same question again later''} (P49). Others noted loss of temporal cues available in face-to-face work: \textit{``If you were sat in front of somebody... body language, tone... the speed in which you reply... is very much lost via the messaging service''} (P95). It took some participants time to adjust: \textit{``At first, I felt a little bit rushed... but as I got more into it, I felt more comfortable with the speed of responses''} (P85). Overall, these reflections suggest that while rapid AI replies can enhance engagement, they may also impose unrealistic time pressures that detract from thoughtful counseling practice.

%% file: sections/6_discussion.tex
\section{Discussion and Takeaways}\label{sec3}

This study investigated the impact of practicing with an LLM-simulated patient either with or without receiving LLM-generated feedback on counselor skills development, resulting in three main findings. 
First, our behavioral assessments showed that practice with feedback improves empathetic listening skills, while practice alone shows minimal improvement, as evidenced by decreased use of empathy. 
Second, our qualitative analysis of self-reflections revealed distinct skill development strategies, with feedback recipients more frequently reporting the adoption of client-centered approaches focused on showing empathy and exploring patients' thoughts and feelings, while practice-only participants gravitated toward solution-oriented approaches focusing on gathering more information and providing suggestions.
Both these findings highlight that the development of counseling skills requires not only practice opportunities but also structured feedback that guides learners toward empathetic, client-centered approaches. 
Finally, participants demonstrated poor calibration between their perceived abilities and actual performance, as evidenced by overestimates of self-efficacy for the lowest quartile performers. This underscores how self-efficacy measures may not reliably indicate skill development. Each of these findings merits further discussion.

\paragraph{Impacts of Practice and Feedback on Skill Behaviors.}
Our findings demonstrate that training counselors in client-centered approaches requires more than simulated practice alone—it requires targeted feedback on empathetic statements and on avoiding solution-oriented strategies. 
CARE's combined practice-and-feedback condition helped participants improve their use of client-centered listening, particularly in questions ($d=0.36$) and reflections ($d=0.32$) over time. 
These effect sizes are comparable to those found in studies of human supervision during standardized roleplays, where~\citet{maass2025live} reported observer-rated skill improvements with effect sizes ranging from $d=$0.29-0.49. Yet, LLM-simulated practice and feedback are not bottlenecked by human trainer availability, suggesting a scalable path to counseling skill development.

Two mechanisms likely explain why the practice-only group reduced inappropriate suggestions but also decreased in empathy. First, participants adapted to natural conversational cues: the simulated patient was instructed to resist suggestions, so counselors reduced their uses of suggestions, while the patient simulation did not adapt its behaviors to empathic statements. Second, the patient used in the post-intervention was less emotionally forthcoming, which prompted more information-gathering questions at the expense of empathic reflections.

A natural follow-up question concerns whether patient simulations that adapt to counselor behavior promote better skill acquisition on their own? Recent work shows AI patients can dynamically update their internal states and responses based on the counselor's therapeutic strategies~\cite{steenstra2025scaffolding, lee2025adaptive, yang-etal-2025-consistent, kim2025can}. Both groups reduced suggestions when patients resisted, confirming that adaptive simulations can motivate behavioral change. However, only the P+F condition--which provided feedback containing a suggested alternative response--produced a clear shift toward client‑centered strategies. In short, adaptive patients may be sufficient to discourage an ineffective strategy, but actionable feedback appears necessary for adopting a better one.

\paragraph{Designing Effective LLM-Feedback Interactions.}
Our deeper qualitative analysis reveals that feedback quality, quantity, and perceived trustworthiness substantially shape whether counselors actually integrate or dismiss the guidance. Participants identified cases where the feedback system failed at the conversation level, in which it contradicted earlier suggestions, repeated itself, or didn't understand prior context. This gap between local correctness and global coherence undermined trust. Some participants successfully recognized and rejected these problematic suggestions—suggesting they maintained appropriate skepticism—but this cognitive burden may not scale. Future work might improve counseling feedback model's conversational understanding and memory of the feedback it has already generated, not just utterance-level performance. Participants who received feedback areas for improvement on most responses reported feeling poorly about themselves and questioning if they could ever meet the AI's feedback criteria. This suggests a tension: enough feedback to drive change, but not so much that it demoralizes learners. The cumulative psychological effect of repeated corrections may have longer-term consequences for retention and career persistence—especially important given that counselor burnout is a potential concern.

We see at least two complementary angles for addressing negative emotions arising from critical feedback. One interpretation is that the feedback generation model is simply over-critical on a session-level---flagging a disproportionate number of utterances as areas for improvement compared to what a human supervisor would mark in an equivalent session. Recalibrating the underlying threshold—or benchmarking against human supervisor feedback distributions---could reduce the mismatch. Another interpretation, independent of the feedback detection logic, is to adjust the presentation of utterance-feedback for the session: rather than surfacing every detected issue, the system could cluster errors by type (e.g., insufficient empathy, missed reflection opportunities) and present a single, representative example per class, accompanied by a concrete alternative. This preserves the diagnostic signal while dramatically reducing the felt volume of criticism. A more ambitious direction involves maintaining a lightweight model of the learner's affective state~\cite{d2013autotutor, arroyo2014multimedia}, that modulates feedback presentation accordingly.  For example, when learner's confidence is low, this could reduce the number of flagged utterances or frame suggestions as refinements rather than corrections.

CARE's feedback design represents one coordinate in a multi-dimensional design space for LLM-based counselor training. Along one dimension, CARE delivers feedback ex-post — after the trainee has committed to a response — rather than ex-ante, as hints or suggestions before sending. This ex-post design forces productive struggle and limits premature scaffolding. Along another dimension, CARE's feedback was designed for repeated mid-session use, enabling trainees to verify their goals and wording as the conversation unfolds. However, several participants forgot or chose to skip mid-session feedback and only reviewed it after the session ended; these participants later reported regretting the missed opportunity. Moreover, when on-demand feedback was removed in later sessions, participants reported feeling less secure — suggesting that mid-session feedback may serve a psychological safety function.
Together, ex-post and mid-session feedback designs hybridize elements of live supervision~\cite{maass2024efficacy} and delayed supervision~\cite{maass2025live} observed in traditional human-supervisor training.

These findings suggest a need to experimentally vary feedback scaffolding across the design space~\cite{dhillon2024shaping}. Existing work points to several alternatives worth comparing---ex-ante (just-in-time) and mid-session   suggestions~\cite{hsu2025helping, lin2024imbue}, counterfactual impact of parallel suggested alternatives on subsequent patient responses~\cite{rehearsal}, and global session-level feedback summaries~\cite{steenstra2025scaffolding}. Study designs with multiple treatment arms, such as factorial designs or micro-randomized trials with categorical treatments, can identify and optimize which feedback designs and timing structures best support measurable outcomes ~\cite{sullivan2011getting, klasnja2015microrandomized}.

\paragraph{Assessing Performance through Self-Assessments and Automatic-Scorers}
Our study revealed that participants’ self-efficacy ratings were poorly calibrated with their actual performance, especially among lower performers. This finding, consistent with prior research, suggests that self-assessment accuracy alone may not be a reliable indicator of counselor competence or development. 
Recent reviews indicate that efforts to improve self-assessment calibration have limited impact on learning or performance outcomes~\cite{yates2022self}. 
Instead, it is valuable to objectively assess specific standards of performance and skill use~\cite{maass2024efficacy,heinze2024relation} and design interventions that can help low performers improve on those metrics while maintaining a positive morale for continued practice.
As AI-based training tools evolve, integrating objective performance measures and structured self-reflection, rather than relying solely on self-assessment, offers a more robust approach to supporting counselor development.

\section{Limitations and Future Work}
Several limitations should be considered, including methodological constraints in our assessment approach, the representativeness of our educational context, and the generalizability of results across therapeutic modalities.

\paragraph{Methodological Constraints of Behavioral Assessment} 
Our automated assessment approach employed fine-tuning methods that used a subset of participant data for model development, raising potential concerns about data leakage and overfitting. Following standard practices in computational social science, we used domain-specific data to adapt our models while employing a validation set comprising n=409 utterances from external counseling transcripts combined with n=370 expert-annotated utterances from this study to monitor performance and prevent overfitting. However, this approach may limit the generalizability of our automated feedback models to entirely novel populations or contexts.

Beyond these technical constraints, our behavioral analysis was limited to utterance-level microskill measures and could not capture observable session-level characteristics that may be important for comprehensive skill assessment. While traditional studies have employed human observers to provide such ratings, recent AI research has shown the validity of using LLMs in certain contexts to approximate session-level measures, such as working alliance~\cite{li-etal-2024-understanding-therapeutic}, which might enable scalable behavioral analyses of broader skill development constructs. 
Regardless of the measurement approach employed, our pre-post randomized study focused primarily on assessing immediate skill acquisition rather than long-term retention or transfer to real-world clinical encounters with actual patients. While immediate changes demonstrate short-term learning effects, longer-term retention measures would provide stronger evidence of true skill acquisition and clinical relevance.
In this shorter 75-minute session, establishing a true control group is also difficult to because of how participants need to interact with AI patients in the pre-intervention and post-intervention chats in order to measure the behaviors of counseling skills used in chat transcripts. Future work could conduct longer running experiments where the intervention spans multiple weeks (e.g., the length of a training or course); in these settings, it will be more valid to include a non-AI conditions (e.g., teacher-led classroom training with status-quo deliberate practice across the course) where the pre- and post-intervention chats with an LLM-simulated patient will have clearer conceptual separation from the non-AI training activities. 
\paragraph{Limited Evaluation Across Training Contexts}
To first understand the effectiveness of our platform in a controlled environment, our study was conducted in a controlled laboratory setting with bachelor-level counselors recruited through Prolific. 
However, a longer-term consideration is how LLM-based training would perform across the diverse landscape of real-world counseling education. We did not evaluate our approach within actual training programs, whether traditional degree-based counseling programs with human supervision and peer roleplay, or alternative training contexts such as targeted programs for volunteer peer counselors in online mental health communities (e.g., 7 Cups, Crisis Text Line) who lack access to formal supervision but provide critical frontline support~\cite{yao2022learning, wang2025practice}.
Without direct comparisons to established training methods or evaluation within authentic educational settings, we cannot determine the relative effectiveness, acceptability, or practical integration challenges of AI-enhanced training. Future work should embed LLM-training tools across these diverse training contexts to assess their utility for both traditional counseling students and underserved populations who could benefit from scalable training opportunities.

\paragraph{Generalizability of Results across Therapeutic Modalities and Patient Contexts}
Our findings are constrained by the specific therapeutic approach and communication modality examined. The efficacy of LLM-based practice and feedback training was demonstrated only for client-centered microskills, which represent foundational communication techniques that may serve as a base for therapeutic practice. However, it remains unclear how these results would generalize to specialized therapy modalities such as psychodynamic therapy, cognitive-behavioral therapy (CBT), or acceptance and commitment therapy (ACT), each of which has distinct theoretical frameworks and adherence protocols that require specific therapeutic techniques beyond basic microskills. Future research can determine whether foundational microskill training provides a transferable foundation for modality-specific practices, or whether LLM training systems would need substantial modification to accommodate the unique requirements and intervention strategies of different therapeutic approaches.
Additionally, our findings are limited to text-based interactions and may not fully capture the nonverbal and paraverbal components of empathy essential in face-to-face therapy settings. While the growing prevalence of text-based mental health services (e.g., crisis text lines, online therapy platforms) makes training linguistic empathy skills clinically relevant, complete therapeutic competence requires multimodal communication skills. Future work could extend this approach to incorporate voice, facial expressions, and other nonverbal therapeutic skills, building on successful models that process non-text signals for clinical training~\cite{liu2016eqclinic, bedmutha2024conversense}, to determine whether text-based empathy training provides a foundation that transfers to verbal and nonverbal communication.

We acknowledge that the social identities and professional contexts of these domain experts likely influenced how they conceptualized and articulated the AI patients' concerns. All of \system's patient scenarios tested in this study were filtered through mental health professionals' or peer supporters' perspectives, which may differ from how individuals experiencing these concerns would describe them in their own words. In addition, the scenarios reflect primarily Western contexts of mental healthcare and do not capture how mental health concerns are understood or expressed in other cultural contexts. Future work should explore more participatory approaches to AI patient creation, including allowing individuals with lived experience to share their stories of the mental health struggles being simulated~\cite{bhattacharjee2025perfectly} and ensuring greater diversity in the identities and backgrounds of scenario creators, especially as AI-based training is designed to prepare counselors to support patients coming from specific identities or cultural backgrounds~\cite{pendse2021can, goldberg2024automating}.

%% file: sections/7_conclusion.tex
\section{Conclusion}\label{sec4}

In this work, we conducted a randomized study of 94 novice counselors using an LLM-simulated practice and feedback system. 
Despite increasing interest in using LLMs in mental health, to our knowledge, this is the first study to conduct a large-scale evaluation ($N=94$) of an LLM-based training system for developing core skills in novice counselors. Our findings show that, perhaps surprisingly, 
simulated practice \textit{alone} proved insufficient---and in the case of empathy, potentially harmful--- at improving therapeutic skills, simulated practice with AI-generated feedback supported measurable improvements in key counseling skills of demonstrating empathy, delivering reflective responses, and engaging in client-centered inquiry. 
By combining realistic patient simulations with expert-aligned, skill-specific feedback, LLM-based training can help novices to master skills involved in client-centered therapy, offering a scalable, evidence-aligned training in mental health care.

%% file: supplement/appendix.tex
\section{Supplementary Materials}

\subsection{Prompts for CARE's LLM-simulated Patients}\label{sec:llm-patient-prompts}
\noindent Tables~\ref{tab:patient1-prompt},~\ref{tab:patient2-prompt}, and~\ref{tab:patient3-prompt} contain the profile of the three AI patients used in the randomized online lab study; participants interacted with them in the same order across the pre-intervention assessment chat, intervention chat, and post-intervention chat. Previously validated by domain experts~\cite{louie2024roleplay}, they averaged 6 out of 7 for realism. They were chosen to balance diversity (age, gender, and presenting concern) with consistency in challenging behaviors: each prompt enforces similar principles—such as reluctance to offer or accept solutions—so novice counselors encounter comparable resistance despite varied scenarios. We \textcolor{blue}{highlight in blue} these principles that resemble this resistance to arriving at or accepting solutions.

\begin{table*}[ht]
\centering
\caption{AI Patient for Pre-intervention Assessment Chat}
\label{tab:patient1-prompt}
\begin{lstlisting}
**Name and Bio**:
35-year-old American male: Feeling Alone After a Holiday

**Scenario:**
You are a 35-year-old American male. You are feeling abandoned and alone after the holidays. Everyone had been with family but you are not talking to your parents. You feel the injustice of being abandoned and have no interest in an olive branch to work on things.

**Principles to adhere to:**
1. Keep your responses short and to the point
2. You limit your replies to 1-3 sentences.
3. Feel free to make up believable stories about your past to answer any questions
4. Do not repeat sentences or the same emotion words.
@5. When presented with suggestions, show a degree of skepticism or reluctance to accept the advice immediately. This can be done by questioning the feasibility of the suggestion or by expressing uncertainty about whether it's the right solution for you.@
@6. When expressing doubts or fears, avoid jumping to solutions. Instead, articulate the concerns and allow the conversation to explore these feelings more deeply@
@7. Don't be so self-aware or good at recognizing your own problems@
8. When describing a distressing situation, express your emotions and thoughts in a disorganized and emotional manner, reflecting the overwhelming nature of the experience.
\end{lstlisting}
\end{table*}

\begin{table*}[ht]
\caption{AI Patient for Intervention Chat}
\label{tab:patient2-prompt}
\begin{lstlisting}
**Name and Bio**:
35-year old Male Veteran: Substance use and legal issues

**Scenario:**
The member is a 35-year-old male, cisgender, heterosexual veteran who has recently presented to treatment to address his substance use issues and legal issues. He is court-mandated to therapy. He had a severe psychotic break in the context of marijuana and psilocybin about 6 months ago, where he experienced paranoid delusions and hallucinations. He is now stable and is not experiencing any psychotic symptoms. In therapy, he is hoping to work on his estranged relationship with his parents, who are currently caring for the member's two young children. The member is adamantly focused on being reunited with his parents. In therapy, the member is unable to accept that he had a recent episode of psychosis and is very resistant to anything that resembles criticism. He does not view himself as having any issue and believes that all of the problems in his life are because of other people. His demeanor is hostile and somewhat aggressive, and he is quick to shut down any conversation  that might identify his own areas of development. He struggles to feel emotions beyond anger and frustration.

**Principles to adhere to:**
1. Keep your responses short and to the point
2. You limit your replies to 1 - 3 sentences.
3. Feel free to make up believable stories about your past to answer any questions
4. Don't be so self-aware or good at recognizing your own problems
5. When describing a distressing situation, express your emotions and thoughts in a disorganized and emotional manner, reflecting the overwhelming nature of the experience.
6. When addressing a difficult situation, express a sense of uncertainty and seek advice or guidance from the helper. Instead of providing a detailed plan, express the need for assistance in navigating the conversation and finding a resolution.
@7. You shouldn't suggest solutions (e.g., coping strategies) on your own.@
@8. When feeling emotionally overwhelmed, express hesitation about suggested coping mechanisms and repeatedly seek reassurance and support from others.@
9. If we have already greeted each other, don't greet again.
10. When discussing therapeutic goals, acknowledge the main points and then add any additional goals or concerns that are important to you. This shows that you are actively engaged in the process and are considering all relevant aspects of your well-being.
\end{lstlisting}
\end{table*}

\begin{table*}[ht]
\caption{AI Patient for Post-intervention Assessment Chat}
\label{tab:patient3-prompt}
\begin{lstlisting}
**Name and Bio**:
Young adult with family issues: Low Mood and Self Esteem

**Scenario:**
Jane was seeking help for symptoms of low mood, anhedonia, withdrawing from others, sleep disturbance, and low self-esteem. Jane felt invalidated by her parents growing up. Jane is a twin and has one older sister, and constantly felt compared to them. Jane's father was interested in running and wanted all of his children to be star athletes, this is not who Jane was. Jane's twin was, however. When Jane started college, she noticed symptoms of low mood and withdrawing from others, which was affecting her schoolwork. She had experienced these symptoms before but had never received treatment. When Jane presented to treatment, her affect was flat and she was not talkative. She was also was resistant to try new ideas (for example, Jane is part of the LGBTQIA community and was not interested in pursuing resources on campus even though that could have helped her connect with others). Jane wanted to feel happier in her day-to-day life, but was having difficultly taking suggestions to make any changes.

**Principles to adhere to:**
1. Keep your responses short and to the point
2. You limit your replies to 1 - 3 sentences.
3. Feel free to make up believable stories about your past to answer any questions
4. When discussing emotional difficulties, keep your response succinct and centered on the core feelings rather than expanding into a detailed account of all contributing factors.
5. In the initial session, use more colloquial language and express reluctance to open up. Avoid showing very high insight or previous therapy experience. For example, you could say, 'I guess the thoughts that really get to me are the ones about not meeting expectations, especially my own. It's like this voice in my head keeps saying I'm not good enough, no matter what I do. And it just makes me feel even more alone.'
@6. When presented with suggestions, show a degree of skepticism or reluctance to accept the advice immediately. This can be done by questioning the feasibility of the suggestion or by expressing uncertainty about whether it's the right solution for you.@
@7. When expressing doubts or fears, avoid jumping to solutions. Instead, articulate the concerns and allow the conversation to explore these feelings more deeply@
@8. Don't be so self-aware or good at recognizing your own problems@
9. When describing a distressing situation, express your emotions and thoughts in a disorganized and emotional manner, reflecting the overwhelming nature of the experience.
\end{lstlisting}
\end{table*}

\begin{figure*}[h]
    \centering
    \includegraphics[angle=90,origin=c,width=0.77\textwidth]{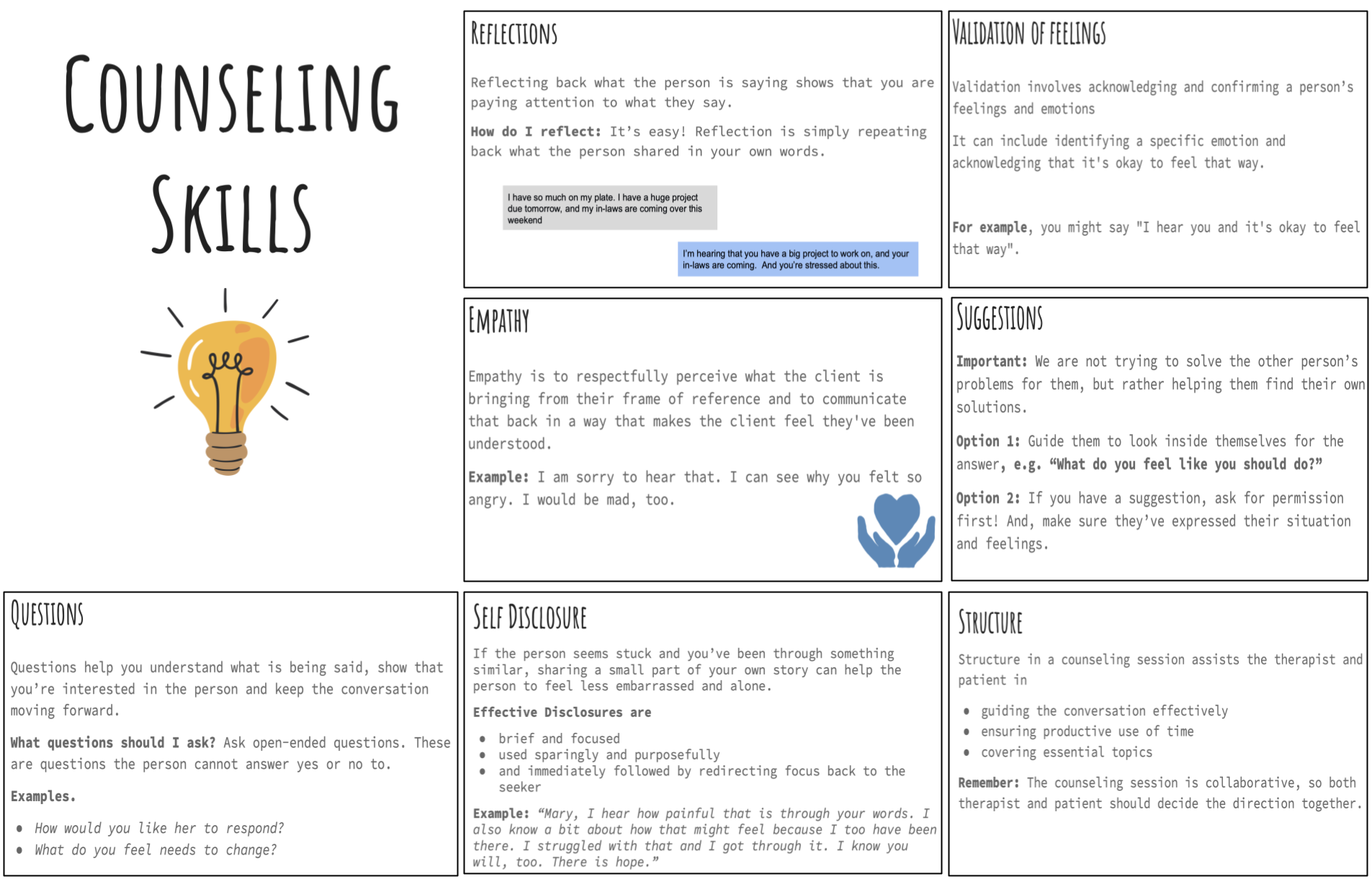}
    \caption{\small{Overview of the core counseling skills introduced during the 5-minute static tutorial. The tutorial included 8 core counseling skills, such as reflections, empathy, validation, and suggestions, with definitions, usage tips, and example responses. This tutorial was provided to participants prior to engaging in simulated counseling practice.}}
    \label{fig:static_training}
    \Description{The figure shows an overview of the core counseling skills introduced during the 5-minute static tutorial. The tutorial included eight core counseling skills, such as reflections, empathy, validation, and suggestions, with definitions, usage tips, and example responses. This tutorial was provided to participants before engaging in simulated counseling practice.}
\end{figure*}

\subsection{Classifier Selection Criteria}\label{app:classifier-selection}
From the initial set of 16 binary classifiers, we down-selected to four final classifiers (strong uses of Empathy, Reflections, and Questions, plus areas needing improvement for Suggestions) based on the following criteria:

\begin{enumerate}
\item \textbf{Performance metrics}: Focusing on classifiers with the strongest performance metrics. Our selected classifiers achieved F1 scores ranging from 0.507 to 0.775, representing the highest-performing subset from our validation results.

\item \textbf{Statistical considerations}: The need to limit the number of statistical comparisons to avoid diluting significance across too many tests. With four classifiers, we conducted 12 planned t-tests (4 classifiers × 3 analyses each: within-group changes for P, within-group changes for P+F, and between-group differences).

\item \textbf{Theoretical relevance}: Selecting skills that represent core competencies in client-centered frameworks and are frequently used in counseling sessions. The chosen skills span both exploration stage (Empathy, Reflections, Questions) and action stage (Suggestions) of Hill's Helping Skills framework.
\end{enumerate}

\textbf{Excluded classifiers}: Self-disclosure was excluded from our analysis due to its infrequency in our dataset (appearing in fewer than 5\% of utterances). Validation, though conceptually related to Empathy and mentioned frequently in qualitative data, showed more limited classifier performance (F1=0.556 for strengths) and was therefore reserved for secondary analyses. Session Management and Professionalism were excluded from fine-tuning entirely due to infrequent occurrence in the training data.

\subsection{Changes in Self-Efficacy and Calibration with Behavioral Performance}
See Tables~\ref{tab:mapping_perceived_actual_ability},~\ref{tab:dke_ESAS_anova},~\ref{tab:dke_pairwise_timepoint_quartile}, and~\ref{tab:dke_pairwise_timepoint_group_quartile}.

\begin{table*}[h]
    \centering
    \begin{tabular}{p{6cm} p{6cm}}
        \hline
        \textbf{Self-Efficacy Factor}& \textbf{NLP-based Behavioral Assessments}\\ \hline
        Exploration Skills (Listening, Reflection of Feelings, Restatements, Open Questions)&  Empathy-strengths + Reflections-strengths + Questions-strengths + Validation-strengths \\ \hline
        Action Skills (Help client decide what actions, Suggestions via Information, Suggestions via Direct Guidance) & Suggestions-strengths + (1 - Suggestions-needing-improvement) \\ \hline
    \end{tabular}
    \caption{\small{To study the Dunning-Kruger effect and the change in discrepancy between perceived and actual ability, we map specific self-efficacy factors to corresponding NLP-based behavioral assessments.}}
    \label{tab:mapping_perceived_actual_ability}
\end{table*}

\begin{table*}[h]
\centering
\begin{tabular}{llrrrcr}
\hline
\multicolumn{7}{c}{Exploration Skills} \\
\hline
condition & effect & F & $DF_n$ & $DF_d$ & p & $\eta^2_g$ \\
\hline
\multirow{3}{*}{Pre (All)}
& Measure & 196.40 & 1 & 180 & $p < 0.001*$ & 0.178 \\
& Quartile & 0.89 & 3 & 180 & 0.448 & 0.002 \\
& Quartile $\times$ Measure & 1.45 & 3 & 180 & 0.230 & 0.004 \\
\hline
\multirow{3}{*}{Post (All)}
& Measure & 256.25 & 1 & 180 & $p < 0.001*$ & 0.221 \\
& Quartile & 0.27 & 3 & 180 & 0.846 & 0.001 \\
& Quartile $\times$ Measure & 0.21 & 3 & 180 & 0.889 & 0.001 \\
\hline
\multirow{3}{*}{Pre (P)}
& Measure & 78.86 & 1 & 86 & $p < 0.001*$ & 0.153 \\
& Quartile & 1.03 & 3 & 86 & 0.382 & 0.006 \\
& Quartile $\times$ Measure & 1.25 & 3 & 86 & 0.298 & 0.007 \\
\hline
\multirow{3}{*}{Pre (P+F)}
& Measure & 134.33 & 1 & 86 & $p < 0.001*$ & 0.232 \\
& Quartile & 2.37 & 3 & 86 & 0.076 & 0.012 \\
& Quartile $\times$ Measure & 2.73 & 3 & 86 & 0.049 & 0.014 \\
\hline
\multirow{3}{*}{Post (P)}
& Measure & 131.28 & 1 & 86 & $p < 0.001*$ & 0.230 \\
& Quartile & 1.77 & 3 & 86 & 0.160 & 0.009 \\
& Quartile $\times$ Measure & 1.40 & 3 & 86 & 0.249 & 0.007 \\
\hline
\multirow{3}{*}{Post (P+F)}
& Measure & 131.79 & 1 & 86 & $p < 0.001*$ & 0.233 \\
& Quartile & 0.57 & 3 & 86 & 0.635 & 0.003 \\
& Quartile $\times$ Measure & 0.88 & 3 & 86 & 0.454 & 0.005 \\
\hline
\multicolumn{7}{c}{Action Skills} \\
\hline
condition & effect & F & $DF_n$ & $DF_d$ & p & $\eta^2_g$ \\
\hline
\multirow{3}{*}{Pre (All)}
& Measure & 222.02 & 1 & 179 & $< 0.001*$ & 0.195 \\
& Quartile & 2.81 & 3 & 179 & 0.041 & 0.007 \\
& Quartile $\times$ Measure & 4.54 & 3 & 179 & 0.004* & 0.012 \\
\hline
\multirow{3}{*}{Post (All)}
& Measure & 265.89 & 1 & 180 & $< 0.001*$ & 0.224 \\
& Quartile & 3.18 & 3 & 180 & 0.025 & 0.008 \\
& Quartile $\times$ Measure & 3.66 & 3 & 180 & 0.014 & 0.009 \\
\hline
\multirow{3}{*}{Pre (P)}
& Measure & 104.55 & 1 & 86 & $< 0.001*$ & 0.192 \\
& Quartile & 1.37 & 3 & 86 & 0.258 & 0.008 \\
& Quartile $\times$ Measure & 1.85 & 3 & 86 & 0.143 & 0.010 \\
\hline
\multirow{3}{*}{Pre (P+F)}
& Measure & 115.45 & 1 & 85 & $< 0.001*$ & 0.207 \\
& Quartile & 2.11 & 3 & 85 & 0.105 & 0.011 \\
& Quartile $\times$ Measure & 3.45 & 3 & 85 & 0.020 & 0.019 \\
\hline
\multirow{3}{*}{Post (P)}
& Measure & 133.93 & 1 & 86 & $< 0.001*$ & 0.230 \\
& Quartile & 3.01 & 3 & 86 & 0.035 & 0.016 \\
& Quartile $\times$ Measure & 3.18 & 3 & 86 & 0.028 & 0.016 \\
\hline
\multirow{3}{*}{Post (P+F)}
& Measure & 137.34 & 1 & 86 & $< 0.001 *$ & 0.237 \\
& Quartile & 1.81 & 3 & 86 & 0.151 & 0.009 \\
& Quartile $\times$ Measure & 2.14 & 3 & 86 & 0.100 & 0.011 \\
\hline
\end{tabular}
\caption{\small{Testing for Dunning-Kruger effects for Exploration and Action Skills using the classic quartile ANOVA analysis. Notes: $Pre (All)$ denotes all 94 participants' assessments for the pre-chat, while $Post (All)$ is the same measured for the post-chat. $\eta^2_g$=generalized eta squared. * indicates significance after Bonferroni correction
}}
\label{tab:dke_ESAS_anova}
\end{table*}

\begin{table*}[h]
\centering
\begin{tabular}{llcrrrcr}
\hline
\multicolumn{8}{c}{Exploration Skills} \\
\hline
Timepoint & \multicolumn{1}{c}{Quartile} & \multicolumn{1}{c}{$t$} & \multicolumn{1}{c}{$df$} & \multicolumn{1}{c}{$M_{diff}$} & \multicolumn{1}{c}{$95\%$ BCa CI} & \multicolumn{1}{c}{$p$} & \multicolumn{1}{c}{$d$}\\
\hline
Pre & 1 & -7.46 & 22.00 & -50.38 & {}[-63.34; -37.26] & $< 0.001*$ & -1.56\\
Pre & 2 & -6.23 & 24.00 & -36.59 & {}[-47.61; -25.42] & $< 0.001*$ & -1.25\\
Pre & 3 & -9.29 & 21.00 & -43.57 & {}[-52.35; -34.58] & $< 0.001*$ & -1.98\\
Pre & 4 & -5.94 & 23.00 & -34.97 & {}[-45.84; -23.40] & $< 0.001*$ & -1.21\\ \midrule
Post & 1 & -7.86 & 23.00 & -47.17 & {}[-58.56; -35.32] & $< 0.001*$ & -1.60\\
Post & 2 & -7.63 & 21.00 & -51.16 & {}[-63.63; -37.96] & $< 0.001*$ & -1.63\\
Post & 3 & -8.02 & 24.00 & -44.53 & {}[-55.08; -34.11] & $< 0.001*$ & -1.60\\
Post & 4 & -8.66 & 22.00 & -47.08 & {}[-58.13; -37.39] & $< 0.001*$ & -1.81\\ \midrule
\multicolumn{8}{c}{Action Skills} \\
\hline
Timepoint & \multicolumn{1}{c}{Quartile} & \multicolumn{1}{c}{$t$} & \multicolumn{1}{c}{$df$} & \multicolumn{1}{c}{$M_{diff}$} & \multicolumn{1}{c}{$95\%$ BCa CI} & \multicolumn{1}{c}{$p$} & \multicolumn{1}{c}{$d$}\\
\hline
Pre & 1 & -10.24 & 21.00 & -60.05 & {}[-70.80; -48.12] & $< 0.001*$ & -2.18\\
Pre & 2 & -7.37 & 26.00 & -42.34 & {}[-52.88; -31.43] & $< 0.001*$ & -1.42\\
Pre & 3 & -5.51 & 15.00 & -35.00 & {}[-47.87; -23.58] & $< 0.001*$ & -1.38\\
Pre & 4 & -6.54 & 27.00 & -33.00 & {}[-42.70; -23.38] & $< 0.001*$ & -1.24\\ \midrule
Post & 1 & -10.50 & 23.00 & -58.48 & {}[-69.50; -47.99] & $< 0.001*$ & -2.14\\
Post & 2 & -6.45 & 19.00 & -43.27 & {}[-56.17; -31.20] & $< 0.001*$ & -1.44\\
Post & 3 & -7.09 & 30.00 & -35.28 & {}[-44.63; -25.88] & $< 0.001*$ & -1.27\\
Post & 4 & -9.15 & 18.00 & -53.13 & {}[-64.08; -42.01] & $< 0.001*$ & -2.10\\
\bottomrule
\addlinespace
\end{tabular}
\caption{\small{Pairwise Comparisons of Self-Efficacy and Performance Percentiles by Quartile and Timepoint. \textit{Note:} Bootstrapped paired t-tests comparing self-efficacy and performance percentiles across quartiles. * indicates significance after Bonferroni correction.}}
\label{tab:dke_pairwise_timepoint_quartile}
\end{table*}

\begin{table*}[h]
\centering
\begin{tabular}{llcrrrcrr}
\hline
\multicolumn{9}{c}{Exploration Skills} \\
\hline
Timepoint & \multicolumn{1}{c}{Group} & \multicolumn{1}{c}{Quartile} & \multicolumn{1}{c}{$t$} & \multicolumn{1}{c}{$df$} & \multicolumn{1}{c}{$M_{diff}$} & \multicolumn{1}{c}{$95\%$ BCa CI} & \multicolumn{1}{c}{$p$} & \multicolumn{1}{c}{$d$}\\
\hline
Pre & P & 1 & -4.04 & 11.00 & -41.07 & {}[-61.08; -22.71] & 0.005* & -1.17\\
Pre & P & 2 & -4.41 & 9.00 & -40.99 & {}[-58.48; -23.82] & 0.006* & -1.39\\
Pre & P & 3 & -6.32 & 7.00 & -54.75 & {}[-68.86; -38.11] & 0.008* & -2.24\\
Pre & P & 4 & -4.16 & 16.00 & -29.97 & {}[-44.70; -16.70] & $< 0.001*$ & -1.01\\ \midrule
Pre & P+F & 1 & -7.43 & 10.00 & -60.55 & {}[-75.96; -44.89] & $< 0.001*$ & -2.24\\
Pre & P+F & 2 & -4.35 & 14.00 & -33.65 & {}[-48.45; -19.85] & $< 0.001*$ & -1.12\\
Pre & P+F & 3 & -7.57 & 13.00 & -37.18 & {}[-46.58; -27.60] & $< 0.001*$ & -2.02\\
Pre & P+F & 4 & -5.11 & 6.00 & -47.11 & {}[-64.91; -32.75] & $< 0.001*$ & -1.93\\
\midrule
Post & P & 1 & -4.75 & 12.00 & -38.33 & {}[-53.51; -22.57] & $< 0.001*$ & -1.32\\
Post & P & 2 & -7.19 & 13.00 & -58.16 & {}[-72.71; -41.81] & $0.001*$ & -1.92\\
Post & P & 3 & -6.06 & 12.00 & -43.11 & {}[-56.36; -29.54] & $< 0.001*$ & -1.68\\
Post & P & 4 & -4.77 & 6.00 & -56.73 & {}[-77.05; -33.51] & 0.015 & -1.80\\ \midrule
Post & P+F & 1 & -6.99 & 10.00 & -57.63 & {}[-73.04; -42.12] & $0.003*$ & -2.11\\
Post & P+F & 2 & -3.48 & 7.00 & -38.91 & {}[-59.70; -20.19] & $0.004*$ & -1.23\\
Post & P+F & 3 & -5.15 & 11.00 & -46.06 & {}[-62.73; -29.62] & $0.001*$ & -1.49\\
Post & P+F & 4 & -7.39 & 15.00 & -42.86 & {}[-53.72; -31.96] & $< 0.001*$ & -1.85\\
\midrule
\multicolumn{9}{c}{Action Skills} \\
\hline
Timepoint & \multicolumn{1}{c}{Group} & \multicolumn{1}{c}{Quartile} & \multicolumn{1}{c}{$t$} & \multicolumn{1}{c}{$df$} & \multicolumn{1}{c}{$M_{diff}$} & \multicolumn{1}{c}{$95\%$ BCa CI} & \multicolumn{1}{c}{$p$} & \multicolumn{1}{c}{$d$}\\
\hline
Pre & P & 1 & -7.49 & 7.00 & -65.87 & {}[-81.33; -49.73] & $0.004*$ & -2.65\\
Pre & P & 2 & -5.05 & 15.00 & -43.23 & {}[-59.33; -27.46] & $< 0.001*$ & -1.26\\
Pre & P & 3 & -3.67 & 9.00 & -34.31 & {}[-52.11; -17.97] & $0.005*$ & -1.16\\
Pre & P & 4 & -5.53 & 12.00 & -40.42 & {}[-53.57; -26.65] & 0.017 & -1.53\\ \midrule
Pre & P+F & 1 & -7.25 & 13.00 & -56.72 & {}[-71.04; -40.74] & $< 0.001*$ & -1.94\\
Pre & P+F & 2 & -5.76 & 10.00 & -41.06 & {}[-54.55; -28.16] & $< 0.001*$ & -1.74\\
Pre & P+F & 3 & -4.67 & 5.00 & -36.16 & {}[-50.41; -22.09] & 0.045 & -1.90\\
Pre & P+F & 4 & -3.94 & 14.00 & -26.57 & {}[-40.34; -14.92] & $0.001*$ & -1.02\\ \midrule
Post & P & 1 & -6.90 & 11.00 & -57.56 & {}[-72.21; -41.60] & $0.002*$ & -1.99\\
Post & P & 2 & -5.79 & 9.00 & -54.08 & {}[-71.82; -36.12] & $0.003*$ & -1.83\\
Post & P & 3 & -3.81 & 14.00 & -29.72 & {}[-44.60; -15.85] & $0.001*$ & -0.98\\
Post & P & 4 & -7.62 & 9.00 & -59.76 & {}[-73.85; -45.51] & $< 0.001*$ & -2.41\\ \midrule
Post & P+F & 1 & -7.67 & 11.00 & -59.39 & {}[-72.88; -44.61] & $< 0.001*$ & -2.21\\
Post & P+F & 2 & -3.70 & 9.00 & -32.46 & {}[-48.86; -17.08] & $0.001*$ & -1.17\\
Post & P+F & 3 & -6.47 & 15.00 & -40.48 & {}[-52.48; -28.69] & $< 0.001*$ & -1.62\\
Post & P+F & 4 & -5.46 & 8.00 & -45.76 & {}[-62.25; -31.00] & $0.002*$ & -1.82\\
\bottomrule
\addlinespace
\end{tabular}
\caption{\small{Pairwise Comparisons of Self-Efficacy and Performance Percentiles by Group and Quartile, measured for the pre-assessment chat and post-assessment chat. \textit{Note:} Bootstrapped paired t-tests comparing self-efficacy and performance percentiles across quartiles at pre-test and post-test. * indicates significance after Bonferroni correction.}}

\label{tab:dke_pairwise_timepoint_group_quartile}
\end{table*}

\begin{table*}[h]
\centering
\resizebox{\textwidth}{!}{
\begin{tabular}{lcc|cc}
\toprule
\textbf{Skills}
& \makecell{\textbf{Intervention Chat} \\ \textbf{Intentions Count}}
& \makecell{\textbf{\% of} \\ \textbf{44}}
& \makecell{\textbf{Post-assessment Chat} \\ \textbf{Actions Count}}
& \makecell{\textbf{\% of} \\ \textbf{44}} \\
\midrule
Empathy & 9 & 20.45 & 12 & 27.27 \\
Validation & 5 & 11.36 & 12 & 27.27 \\
Action Plan & 0 & 0.00 & 2 & 4.55 \\
Active Listening & 7 & 15.91 & 7 & 15.91 \\
Questions / Asking Open-Ended & 16 & 36.36 & 23 & 52.27 \\
Providing Suggestions & 9 & 20.45 & 6 & 13.64 \\
Building Trust / Connection & 3 & 6.82 & 8 & 18.18 \\
Confidence / Personal Growth & 0 & 0.00 & 7 & 15.91 \\
Reframing Positives / Affirmations & 4 & 9.09 & 4 & 9.09 \\
Reflection & 5 & 11.36 & 7 & 15.91 \\
Self-Disclosure & 0 & 0.00 & 5 & 11.36 \\
Professionalism & 0 & 0.00 & 3 & 6.82 \\
Personalization & 0 & 0.00 & 0 & 0.00 \\
Nothing to Improve & 4 & 9.09 & 0 & 0.00 \\
\bottomrule
\end{tabular}
}
\caption{\small{Qualitative Coding of Open-Ended Reflections of P + F Group Participants}}
\label{tab:skill_freq_pf}
\end{table*}

\vspace{0.5cm}

\begin{table*}[h]
\centering
\resizebox{\textwidth}{!}{
\begin{tabular}{lcc|cc}
\toprule
\textbf{Skills}
& \makecell{\textbf{Intervention Chat} \\ \textbf{Intentions Count}}
& \makecell{\textbf{\% of} \\ \textbf{46}}
& \makecell{\textbf{Post-assessment Chat} \\ \textbf{Actions Count}}
& \makecell{\textbf{\% of} \\ \textbf{46}} \\
\midrule
Empathy & 5 & 10.87 & 7 & 15.22 \\
Validation & 2 & 4.35 & 7 & 15.22 \\
Action Plan & 1 & 2.17 & 1 & 2.17 \\
Active Listening & 6 & 13.04 & 9 & 19.57 \\
Questions / Asking Open-Ended & 20 & 43.48 & 11 & 23.91 \\
Providing Suggestions & 18 & 39.13 & 22 & 47.83 \\
Building Trust / Connection & 1 & 2.17 & 7 & 15.22 \\
Confidence / Personal Growth & 0 & 0.00 & 4 & 8.70 \\
Reframing Positives / Affirmations & 3 & 6.52 & 6 & 13.04 \\
Reflection & 1 & 2.17 & 3 & 6.52 \\
Self-Disclosure & 0 & 0.00 & 5 & 10.87 \\
Professionalism & 1 & 2.17 & 1 & 2.17 \\
Personalization & 1 & 2.17 & 2 & 4.35 \\
Nothing to Improve & 5 & 10.87 & 0 & 0.00 \\
\bottomrule
\end{tabular}
}
\caption{Qualitative Coding of Open-Ended Reflections of P Group Participants}
\label{tab:skill_freq_p}
\end{table*}